\documentclass[aps,amsmath,amssymb,twocolumn,floatfix,pra,footinbib,superscriptaddress,longbibliography]{revtex4-1}

\usepackage{graphicx}
\usepackage{epstopdf}

\usepackage[applemac]{inputenc}
\usepackage[T1]{fontenc}
\usepackage{lmodern}
\usepackage[english]{babel}
\usepackage{ae}
\usepackage{units}
\usepackage{color}
\usepackage{url}

\usepackage{amsmath,amssymb,natbib,bm}
\usepackage{psfrag}
\usepackage{amssymb}
\usepackage{amsthm}
\usepackage{mathrsfs}  

\usepackage{fixltx2e}

\usepackage{slashed}

\usepackage[americaninductors]{circuitikz}
\usepackage{tikz}
\usetikzlibrary{arrows}

\usepackage[colorlinks]{hyperref}
\hypersetup{%
	plainpages=true,
	breaklinks=true,
	hypertexnames=false,
	pageanchor=true,
	colorlinks=true,
	linkcolor={blue},
	citecolor={red},
	urlcolor={blue},
	anchorcolor={black}
}

\usepackage{mleftright} 

\newcommand{\ket}[1]{\left|#1\right\rangle}

\newcommand{\be}{\begin{equation}}
\newcommand{\ee}{\end{equation}}
\newcommand{\bea}{\begin{eqnarray}}
\newcommand{\eea}{\end{eqnarray}}

\newcommand{\rd}{\ensuremath{\mathrm{d}}}
\newcommand{\id}{\ensuremath{\,\rd}}

\newcommand{\ketbra}[2]{\left| #1 \rangle \langle #2 \right|}
\newcommand{\brakket}[3]{\mleft\langle #1\mleft| #2 \mright| #3\mright\rangle}

\newcommand{\tr}[1]{\text{Tr}\mleft( #1 \mright)}

\newcommand{\abs}[1]{\mleft|#1\mright|}
\newcommand{\abssq}[1]{\mleft| #1 \mright|^2}

\newcommand{\sz}{\sigma_z}
\newcommand{\sx}{\sigma_x}
\newcommand{\sy}{\sigma_y}

\newcommand{\nn}{\nonumber}

\newcommand{\figref}[1]{\mbox{Fig.~\ref{#1}}}
\newcommand{\tabref}[1]{\mbox{Table~\ref{#1}}}
\newcommand{\secref}[1]{\mbox{Section~\ref{#1}}}

\newcommand{\appref}[1]{\mbox{Appendix~\ref{#1}}}
\renewcommand{\eqref}[1]{\mbox{Eq.~(\ref{#1})}}

\newcommand{\figpanel}[2]{Fig.~\hyperref[#1]{\ref*{#1}(#2)}}
\newcommand{\figpanels}[3]{Fig.~\hyperref[#1]{\ref*{#1}(#2)-(#3)}}
\newcommand{\figpanelNoPrefix}[2]{\hyperref[#1]{\ref*{#1}(#2)}}

\AtBeginDocument{%
    \newwrite\bibnotes
    \def\bibnotesext{Notes.bib}
    \immediate\openout\bibnotes=\jobname\bibnotesext
    \immediate\write\bibnotes{@CONTROL{REVTEX41Control}}
    \immediate\write\bibnotes{@CONTROL{%
    apsrev41Control,author="08",editor="1",pages="0",title="0",year="1"}}
     \if@filesw
     \immediate\write\@auxout{\string\citation{apsrev41Control}}%
    \fi
}%

\begin{document}
\title{Fast multi-qubit gates through simultaneous two-qubit gates}

\author{Xiu Gu}
\email{guxiu1@gmail.com}
\altaffiliation[Present address: ]{Tencent.}
\affiliation{Department of Microtechnology and Nanoscience, Chalmers University of Technology, 412 96 Gothenburg, Sweden}
\affiliation{Tencent Quantum Laboratory, Tencent, Shenzhen, Guangdong 518057, China}

\author{Jorge Fern\'andez-Pend\'as}
\affiliation{Department of Microtechnology and Nanoscience, Chalmers University of Technology, 412 96 Gothenburg, Sweden}

\author{Pontus Vikst\aa{}l}
\affiliation{Department of Microtechnology and Nanoscience, Chalmers University of Technology, 412 96 Gothenburg, Sweden}

\author{Tahereh Abad}
\affiliation{Department of Microtechnology and Nanoscience, Chalmers University of Technology, 412 96 Gothenburg, Sweden}

\author{Christopher Warren}
\affiliation{Department of Microtechnology and Nanoscience, Chalmers University of Technology, 412 96 Gothenburg, Sweden}

\author{Andreas Bengtsson}
\affiliation{Department of Microtechnology and Nanoscience, Chalmers University of Technology, 412 96 Gothenburg, Sweden}

\author{Giovanna Tancredi}
\affiliation{Department of Microtechnology and Nanoscience, Chalmers University of Technology, 412 96 Gothenburg, Sweden}

\author{Vitaly Shumeiko}
\affiliation{Department of Microtechnology and Nanoscience, Chalmers University of Technology, 412 96 Gothenburg, Sweden}

\author{Jonas Bylander}
\affiliation{Department of Microtechnology and Nanoscience, Chalmers University of Technology, 412 96 Gothenburg, Sweden}

\author{G\"oran Johansson}
\affiliation{Department of Microtechnology and Nanoscience, Chalmers University of Technology, 412 96 Gothenburg, Sweden}

\author{Anton Frisk Kockum}
\email{anton.frisk.kockum@chalmers.se}
\affiliation{Department of Microtechnology and Nanoscience, Chalmers University of Technology, 412 96 Gothenburg, Sweden}

\date{\today}

\begin{abstract}

Near-term quantum computers are limited by the decoherence of qubits to only being able to run low-depth quantum circuits with acceptable fidelity. This severely restricts what quantum algorithms can be compiled and implemented on such devices. One way to overcome these limitations is to expand the available gate set from single- and two-qubit gates to multi-qubit gates, which entangle three or more qubits in a single step. Here, we show that such multi-qubit gates can be realized by the simultaneous application of multiple two-qubit gates to a group of qubits where at least one qubit is involved in two or more of the two-qubit gates. Multi-qubit gates implemented in this way are as fast as, or sometimes even faster than, the constituent two-qubit gates. Furthermore, these multi-qubit gates do not require any modification of the quantum processor, but are ready to be used in current quantum-computing platforms. We demonstrate this idea for two specific cases: simultaneous controlled-Z gates and simultaneous iSWAP gates. We show how the resulting multi-qubit gates relate to other well-known multi-qubit gates and demonstrate through numerical simulations that they would work well in available quantum hardware, reaching gate fidelities well above \unit[99]{\%}. We also present schemes for using these simultaneous two-qubit gates to swiftly create large entangled states like Dicke and Greenberg-Horne-Zeilinger states.

\end{abstract}

\maketitle

\tableofcontents


\section{Introduction}

Quantum computers~\cite{Feynman1982, Nielsen2000} hold a promise of eventually being able to tackle complex problems in chemistry~\cite{Cao2019, McArdle2020}, materials science~\cite{Bauer2020}, finance~\cite{Orus2019, Egger2020}, simulation of quantum systems~\cite{Georgescu2014}, and many other fields~\cite{Montanaro2016, Wendin2017, Preskill2018, Cerezo2020}. However, current~\cite{Arute2019, Pino2021, Mooney2021, Blinov2021, Wu2021} and near-term quantum computers are noisy intermediate-scale quantum (NISQ)~\cite{Preskill2018} devices, where decoherence leads to loss of entanglement and coherence among the qubits in the quantum computer after a relatively short time. Thus, such devices can only run quantum circuits with a low depth, i.e., consisting of a low number of sequential quantum gates.

All quantum algorithms can be decomposed into a sequence of universal single- and two-qubit gates~\cite{Barenco1995, Nielsen2000}. Current quantum computers are usually able to implement a universal gate set with arbitrary single-qubit rotations and one or two entangling two-qubit gates. However, many quantum algorithms, e.g., for optimization problems or quantum simulations, require the creation of large-scale entanglement or many-body interactions. Such interactions between three or more qubits result in a large overhead in terms of circuit depth if they are to be decomposed into and compiled from two-qubit gates~\cite{Nielsen2000}. For example, decomposing the three-qubit Fredkin gate requires at least five two-qubit gates~\cite{Smolin1996}.

Motivated by these limitations of NISQ devices, there has recently been several proposals~\cite{Isenhower2011, Zahedinejad2016, Barnes2017, Liebermann2017, Liu2018, Baekkegaard2019, Gullans2019, Daraeizadeh2020, Khazali2020, Loft2020, Rasmussen2020, Rasmussen2020a, Espinoza2020, Young2020, Yu2020, Zhao2020, He2020, Bahnsen2021} for and some implementations~\cite{Monz2009, Reed2012, Patel2016, Levine2019, Roy2020, Ru2021} of multi-qubit gates, as well as proposals for realizing many-body interactions~\cite{Mezzacapo2014, Kafri2017, Chancellor2017, Kounalakis2019, Pedersen2019, Schondorf2019, Petiziol2020}, without having to decompose them into two-qubit gates. However, in general, these proposals and implementations require one or more of the following: additional resources (e.g., extra qubits, modes, energy levels, or initial entanglement); specific complicated connectivity between qubits; setups or components that go beyond what is needed for implementing single- and two-qubit gates; complicated pulse shapes (which, in superconducting circuits, can be distorted due to the response function of drive lines and limited time resolution of arbitrary waveform generators~\cite{Rol2020}); or phenomena specific to a particular quantum-computing platform.

In this article, we show how various multi-qubit gates can be constructed by simply applying multiple two-qubit gates simultaneously to several qubits such that at least one of the qubits is involved in two or more of the two-qubit gates. The multi-qubit gates we propose can thus be implemented in existing quantum hardware adapted to standard single- and two-qubit gates, without any additional components, complicated pulse shapes, or changes in hardware design being required. Furthermore, our multi-qubit gates are as fast as, or faster than, the two-qubit gates from which they are constructed. Although our examples and discussion of experimental feasibility focus on implementations of quantum computing in superconducting circuits~\cite{Wendin2017, Gu2017, Arute2019, Krantz2019, Kockum2019a, Kjaergaard2020, Blais2020, Mooney2021}, our ideas are applicable to any other quantum-computing platforms that implement two-qubit gates in similar ways. The results presented here thus open up avenues for speeding up quantum computation across many different algorithms and systems.

We illustrate our general idea with two specific examples, simultaneous controlled-Z (CZ) gates and simultaneous iSWAP gates, but note that the simultaneous application of other gates also should be explored. In the first example, we consider CZ gates created by activating the transition between states $\ket{11}$ and $\ket{02}$ (or $\ket{20}$)~\cite{Strauch2003, DiCarlo2009}, where $\ket{0}$ is the ground state and $\ket{1}$ is the first excited state of a qubit, and $\ket{2}$ is the second excited state, which typically is outside the computational subspace. Activating the interaction required for two such gates simultaneously to the nearest neighbours in a linear chain of three qubits, with the middle qubit being the one where the second excited state $\ket{2}$ is populated during the gates, results in a three-qubit gate where both CZ and SWAP are applied to the outer qubits conditioned on the middle qubit being in $\ket{1}$. This three-qubit gate, which we denote CCZS~\cite{Schuch2003}, takes \textit{less} time than a \textit{single} CZ and would require at least \text{three} sequential two-qubit gates if it were to be decomposed. The well-known three-qubit iFredkin gate~\cite{Dallaire-Demers2016} can be realized by adding a two-qubit gate after the CCZS gate. Furthermore, by changing the relative strengths of the constituent CZ gates and their detuning, a whole family of three-qubit gates can be created. These gates can be used to create many-qubit entangled states, e.g., a Greenberger--Horne--Zeilinger (GHZ) state~\cite{Greenberger1989, Bouwmeester1999} in a single step or large Dicke states~\cite{Dicke1954, Haroche2006, Shammah2018} in a few steps, and have applications in phase estimation~\cite{Nielsen2000, Kitaev2002}, Hamiltonian simulation~\cite{Poulin2018, Babbush2018, Low2019}, and swap tests~\cite{Buhrman2001} for quantum machine learning~\cite{Zhao2019}.

In our second example, we consider iSWAP gates created by coupling the states $\ket{01}$ and $\ket{10}$. Just like for the CCZS gate above, simultaneous activation of the interaction for two such gates in a linear chain of three qubits creates a three-qubit gate, which we denote DIV for ``divider'' gate. The DIV gate distributes excitations among all three qubits within the subspaces with fixed excitation number in the computational subspace. Similar to the CCZS gate, the DIV gate is faster than the two-qubit gates created by activating the same interactions and can be used to create both GHZ and large Dicke states. By changing the relative strengths of the constituent iSWAP gates and the gate time, a family of different three-qubit DIV gates can be realized. Since both the DIV gates and the CCZS gates conserve the number of excitations, they may find applications in quantum-chemistry calculations with a fixed number of electrons~\cite{Barkoutsos2018} or in the mixing layer of the quantum alternating operator ansatz~\cite{Hadfield2019} for constrained combinatorial optimisation with conserved Hamming weights~\cite{Streif2021}.

This article is organized as follows. In \secref{sec:SimultaneousCZ}, we present the details for generating a family of multi-qubit gates through simultaneous application of multiple CZ gates. We show how this family of three-qubit gates can be decomposed into a sequence of three two-qubit gates and how the three-qubit gates can be used to implement other well-known three-qubit gates through some additional operation. We then present schemes for rapidly generating large entangled states using our three-qubit gates. Finally, we show, through numerical simulations with parameters from state-of-the-art superconducting quantum-computing platforms, that our three-qubit gates are ready to be implemented with high fidelity and short gate times in currently available quantum hardware. In \secref{sec:SimultaneousiSWAP}, we repeat these steps for simultaneous application of iSWAP gates instead of CZ gates. We conclude in \secref{sec:Conclusion} and give an outlook for future work and applications in \secref{sec:Outlook}. Some further analytical calculations for the simultaneous CZ gates with an additional coupling between the outer qubits in the linear chain are given in \appref{app:CZ}.


\section{Simultaneous controlled-Z gates}
\label{sec:SimultaneousCZ}


\subsection{Setup and gate operation}
\label{sec:SetupSimultaneousCZ}

We here consider simultaneous application of CZ gates that are based on making the states $\ket{11}$ and $\ket{02}$ (or $\ket{20}$) resonant~\cite{Strauch2003, DiCarlo2009}. In such gates, the states $\ket{00}$, $\ket{01}$, and $\ket{10}$ do not couple to other states and remain unchanged while the state $\ket{11}$ acquires a $\pi$ phase shift when its population is transferred to $\ket{02}$ (or $\ket{20}$) and back. In superconducting circuits, this can be achieved either by rapidly tuning the frequencies of the two qubits in and out of the desired resonance~\cite{DiCarlo2009, Bialczak2010, Barends2019, Andersen2019, Negirneac2021, Sung2021, Foxen2020, Xu2021} or by parametric modulation of a coupler connecting the two qubits activating the interaction between $\ket{11}$ and $\ket{02}$ (or $\ket{20}$)~\cite{McKay2016, Roth2017, Bengtsson2020, Ganzhorn2020}. For both methods, high gate fidelities have been demonstrated for short gate times. In Ref.~\cite{Negirneac2021}, a gate fidelity of \unit[99.9]{\%} was reached for a gate time of \unit[60]{ns}.


\subsubsection{Hamiltonians and time evolution}
\label{sec:SimultaneousCZHamiltonians}

\begin{figure}[]
\includegraphics[width=\linewidth]{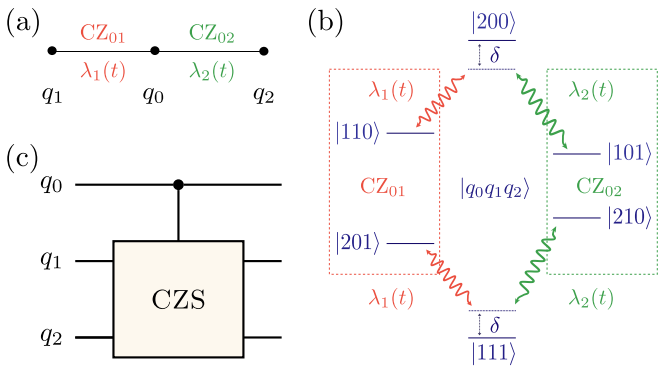}
\caption{Setup and operation for the three-qubit gate realized through simultaneous application of two CZ gates.
(a) The setup considered is a linear chain of three qubits with nearest-neighbour coupling. Going from left to right in the chain, we denote the qubits $q_1$, $q_0$, and $q_2$. The CZ gates $\mathrm{CZ}_{0j}$ between qubits $0$ and $j = \{ 1, 2 \}$ are applied simultaneously by activating a coupling between the $\ket{1_0 1_j}$ and $\ket{2_0 0_j}$ states with the coupling strength $\lambda_j (t)$.
(b) The transitions in the three-qubit system activated by the application of the CZ gates. With the three-qubit states denoted by $\ket{q_0 q_1 q_2}$, the transitions $\ket{11x} \leftrightarrow \ket{20x}$ with $x = \{ 0, 1 \}$ are activated by $\mathrm{CZ}_{01}$ (red), and the transitions $\ket{1x1} \leftrightarrow \ket{2x0}$ are activated by $\mathrm{CZ}_{02}$ (green). We assume that both CZ gate operations are detuned by $\delta$ from resonance.
(c) We denote the three-qubit operation resulting from the simultaneous application of the two CZ gates by CCZS (controlled-CZS), since it applies both CZ and SWAP gates to the target qubits $q_1$ and $q_2$ conditioned on the control qubit $q_0$.
\label{fig:PrincipleCCZS}}
\end{figure}

We first treat the case of three qubits with simultaneous application of two CZ gates (the case of more qubits is discussed further in \secref{sec:CCZSEntanglement} below). We consider the setup shown in \figpanel{fig:PrincipleCCZS}{a}, with the three qubits arranged in a linear chain such that qubit $q_0$ is in the middle, qubit $q_1$ on the left, and qubit $q_2$ on the right. The more complicated case with an additional direct coupling existing between $q_1$ and $q_2$ is discussed in \appref{app:CZ}.

In the setup of \figpanel{fig:PrincipleCCZS}{a}, the transitions between states $\ket{1_0 1_j}$ and $\ket{2_0 0_j}$ are coupled with a strength $\lambda_j (t)$ to implement the standard two-qubit gates $\mathrm{CZ}_{0j}$ between qubits $0$ and $j = \{ 1, 2 \}$ by activating the coupling for a time corresponding to a complete transfer of population from $\ket{1_0 1_j}$ to $\ket{2_0 0_j}$ and back. If both these CZ gates are applied simultaneously, transitions $\ket{110} \leftrightarrow \ket{200} \leftrightarrow \ket{101}$ and $\ket{201} \leftrightarrow \ket{111} \leftrightarrow \ket{210}$, where the states are ordered as $\ket{q_0 q_1 q_2}$, are activated. This creates a $\Lambda$-type three-level system and a $V$-type three-level system, as shown in \figpanel{fig:PrincipleCCZS}{b}.

With all other transitions except the ones shown in \figpanel{fig:PrincipleCCZS}{b} far off resonance, the Hamiltonian for the three-qubit system can be written in the interaction picture as ($\hbar = 1$ throughout this article)
\bea
H &=& \mleft[ \lambda_1 (t) \mleft( \ketbra{110}{200} + \ketbra{111}{201} \mright) \mright. \nn \\
&&\mleft. + \lambda_2 (t) \mleft( \ketbra{101}{200} + \ketbra{111}{210} \mright) + \mathrm {H.c.} \mright] \nn \\
&&+ \delta \mleft( \ketbra{200}{200} - \ketbra{111}{111} \mright),
\eea
where H.c.~denotes Hermitian conjugate and $\delta$ is the detuning, assumed to be the same, for the transitions of both CZ gates. To analyze the time evolution generated by $H$, it is convenient to deal with the two effective three-level systems in \figpanel{fig:PrincipleCCZS}{b} separately.

For the effective $\Lambda$-type three-level system, i.e., the subspace spanned by $\ket{101}$, $\ket{200}$, and $\ket{110}$, we can introduce a new basis: the bright state $\ket{B}$, the dark state $\ket{D}$, and the excited state $\ket{E}$. These states are given by
\bea
\ket{B} &=& e^{i \phi} \sin \frac{\theta}{2} \ket{101} - \cos \frac{\theta}{2} \ket{110}, \\
\ket{D} &=& \cos \frac{\theta}{2} \ket{101} + e^{-i \phi} \sin \frac{\theta}{2} \ket{110}, \\
\ket{E} &=& \ket{200},
\eea
with
\be
\frac{\lambda_2(t)}{\lambda_1(t)} = - e^{i \phi} \tan \frac{\theta}{2}.
\ee
In this basis, the Hamiltonian of this subspace becomes
\bea
H_+ &=& \mleft( \Omega \ketbra{B}{E} + \mathrm{H.c.} \mright) + \delta \ketbra{E}{E} \nn \\
&=& \Omega \sigma_x^{\rm (B, E)} - \frac{\delta}{2} \sigma_z^{\rm (B, E)} + \frac{\delta}{2} I^{\rm (B, E)},
\eea
where
\be
\Omega = \sqrt{\abssq{\lambda_1 (t)} + \abssq{\lambda_2 (t)}}
\ee
and $\sigma_i^{\rm (B, E)}$ are the Pauli matrices in the basis of $\ket{B}$ and $\ket{E}$.

We now consider the simple case where $\lambda_1$, $\lambda_2$, and $\delta$ are time-independent. In that case, the time evolution for the three-level system only affects the two-level subspace spanned by $\ket{B}$ and $\ket{E}$. The time-evolution operator becomes
\bea
&&U^{\rm (B, E)} (t) = e^{- i \frac{\delta t}{2}} \times \\
&&\mleft[ \cos \mleft( t \sqrt{\Omega^2 + \frac{\delta^2}{4}} \mright) - i \sin \mleft( t \sqrt{\Omega^2 + \frac{\delta^2}{4}} \mright)\vec{n}^{\rm (B, E)} \cdot \vec{\sigma}^{\rm (B, E)} \mright], \nn
\label{eq:evolution}
\eea
where
\be
\vec{n}^{\rm (B, E)} = \frac{1}{\sqrt{\Omega^2 + \frac{\delta^2}{4}}} \mleft( \Omega, 0, - \frac{\delta}{2} \mright).
\ee

For this time evolution to yield a useful gate, we need to eliminate any leakage to the state $\ket{E} = \ket{200}$, since it is outside the computational subspace. When starting in the computational subspace, the shortest evolution time which fulfils this condition is
\be
t_{\rm gate} = \frac{\pi}{\sqrt{\Omega^2 + \frac{\delta^2}{4}}}.
\ee
After this time, the states $\ket{B}$ and $\ket{E}$ both acquire a phase factor $- e^{- i \gamma}$, where
\be
\gamma = \frac{\pi \delta}{\sqrt{4 \Omega^2 + \delta^2}},
\ee
while the dark state $\ket{D}$ remains unchanged. Since $\ket{B}$ and $\ket{D}$ also constitute a basis for the subspace spanned by $\ket{101}$ and $\ket{110}$, the effect of the time evolution can be written~\cite{Sjoqvist2016, Sjoqvist2012}
\be
\ketbra{D}{D} - e^{- i \gamma} \ketbra{B}{B} = e^{\frac{i}{2} (\pi - \gamma)} e^{- \frac{i}{2} (\pi - \gamma) \vec{n} \cdot \vec{\sigma}},
\label{eq:2level}
\ee
where
\bea
\vec{n} &=& (\sin \theta \cos \phi, \sin \theta \sin \phi, \cos \theta), \\
\vec{\sigma} &=& (\sx, \sy, \sz).
\eea
Here, the Pauli matrices $\sigma_i$ are in the basis of $\ket{101}$ and $\ket{110}$.

A similar analysis can be performed for the effective $V$-type three-level system, i.e., the subspace spanned by $\ket{111}$, $\ket{210}$, and $\ket{201}$. Introducing the new basis states
\bea
\ket{B'} &=& \sin \frac{\theta}{2} e^{- i \phi} \ket{210} - \cos \frac{\theta}{2} \ket{201}, \\
\ket{D'} &=& \cos \frac{\theta}{2} \ket{210} + \sin \frac{\theta}{2} e^{i \phi} \ket{201}, \\
\ket{E'} &=& \ket{111},
\eea
the Hamiltonian of this subspace can be written as 
\be
H_- = \mleft( \Omega \ketbra{B'}{E'} + \rm{H.c.} \mright) - \delta \ketbra{E'}{E'}.
\ee
Thus, time evolution until the gate time $t_{\rm gate}$ will lead to both $\ket{B'}$ and $\ket{E'}$ acquiring a phase factor $- e^{i \gamma}$. However, $\ket{B'}$ and $\ket{D'}$ span the subspace of the states $\ket{201}$ and $\ket{210}$, neither of which is in the computational subspace, and thus will not be populated in the intial or final states of the gate. The only effect of the gate in the effective $V$-type three-level system is thus to bestow a phase factor $- e^{i \gamma}$ on $\ket{111}$.


\subsubsection{The family of three-qubit gates}

Summarizing the results from the analysis above, we see that the eight states in the computational subspace of the three qubits are affected as follows: $\ket{101}$ and $\ket{110}$ obey the time evolution given by \eqref{eq:2level}, $\ket{111}$ will acquire a phase factor $- e^{i \gamma}$, and all the other states are unchanged. This is similar to the three-qubit Fredkin gate (controlled-SWAP)~\cite{Fredkin1982, Milburn1989, Smolin1996, Nielsen2000}, which swaps the states of two target qubits conditioned on the state of a control qubit, i.e., $\ket{101}$ and $\ket{110}$ are swapped if the first qubit is the control qubit. Our gate, which we denote CCZS [see \figpanel{fig:PrincipleCCZS}{c}], also implements a SWAP-like operation on the outer qubits $q_1$ and $q_2$, conditioned on the middle qubit $q_0$, but adds phase factors to $\ket{101}$, $\ket{110}$, and $\ket{111}$. The gate can be written as
\be
\mathrm{CCZS} (\theta, \phi, \gamma) = \ketbra{0}{0}_0 \otimes \mathbb{I}_1 \otimes \mathbb{I}_2 + \ketbra{1}{1}_0 \otimes U_{\rm CZS} (\theta, \phi, \gamma),
\label{eq:CCZS}
\ee
where
\bea
&&U_{\rm CZS} (\theta, \phi, \gamma) = \label{eq:UCZS} \\
&&\begin{bmatrix}
1 & 0 & 0 & 0 \\
0 & - e^{i \gamma} \sin^2 \frac{\theta}{2} + \cos^2 \frac{\theta}{2} & \frac{1}{2} \mleft( 1 + e^{i \gamma} \mright) e^{- i \phi} \sin \theta & 0 \\
0 & \frac{1}{2} \mleft( 1 + e^{i \gamma} \mright) e^{i \phi} \sin \theta & - e^{i \gamma} \cos^2 \frac{\theta}{2} + \sin^2 \frac{\theta}{2} & 0 \\
0 & 0 & 0 & - e^{i \gamma}
\end{bmatrix} \nn
\eea
and the parameters $\theta$, $\phi$, and $\gamma$ are set by the coupling strengths $\lambda_1$, $\lambda_2$ and the detuning $\delta$ according to the relations
\bea
- e^{i \phi} \tan \frac{\theta}{2} &=& \frac{\lambda_2}{\lambda_1}, \label{eq:ThetaPhiRelation} \\
\gamma &=& \frac{\pi \delta} {\sqrt{4 \Omega^2 + \delta^2}} \in (-\pi, \pi), \\
\Omega &=& \sqrt{\abssq{\lambda_1} + \abssq{\lambda_2}}. \label{eq:parameter}
\eea
%


\subsubsection{Examples of three-qubit gates}
\label{sec:Examples3QubitGatesCCZS}

It is illuminating to study a few of the simplest parameter choices for the CCZS gate. If we set $\lambda_1 = \lambda$, $\lambda_2 = 0$, and $\delta = 0$, we recover the two-qubit CZ gate acting on $q_0$ and $q_1$. In the same way, if instead $\lambda_1 = 0$, $\lambda_2 = \lambda$, and $\delta = 0$, we obtain the two-qubit CZ gate acting on $q_0$ and $q_2$. The gate time for these gates is $t_{\rm gate} = \pi / \lambda$.

If we instead apply both these CZ gates simultaneously, i.e., $\lambda_1 = \lambda_2 = \lambda$ and $\delta = 0$, we obtain $\mathrm{CCZS} (\theta = \pi / 2, \phi = \pi, \gamma = 0)$, for which
\be
U_{\rm CZS} (\pi / 2, \pi, 0) =
\begin{bmatrix}
1 & 0 & 0 & 0 \\
0 & 0 & - 1 & 0 \\
0 & - 1 & 0 & 0 \\
0 & 0 & 0 & - 1
\end{bmatrix}.
\label{eq:UCZSpi2pi0}
\ee
The gate time for this gate is $t_{\rm gate} = \pi / \sqrt{2} \lambda$, i.e., this three-qubit gate is a factor $\sqrt{2}$ faster than the two-qubit CZ gates generated by the two interactions from which the CCZS gate is constructed.

We can set the phase $\phi$ by adjusting the relative phase of the coupling strengths $\lambda_1$ and $\lambda_2$. For $\lambda_1 = \lambda$, $\lambda_2 = - \lambda e^{-i\phi}$, and $\delta = 0$, the controlled part of the gate becomes
\be
U_{\rm CZS} (\pi / 2, \phi, 0) =
\begin{bmatrix}
1 & 0 & 0 & 0 \\
0 & 0 & e^{-i\phi} & 0 \\
0 & e^{i\phi} & 0 & 0 \\
0 & 0 & 0 & - 1
\end{bmatrix}.
\label{eq:UCZS_phi}
\ee
The gate time remains $t_{\rm gate} = \pi / \sqrt{2} \lambda$.

Further tuning can be achieved by changing the relative amplitudes of the coupling strengths $\lambda_1$ and $\lambda_2$. Setting $\lambda_1 = \lambda$, $\lambda_2 = - K \lambda e^{-i\phi}$, and $\delta = 0$, we have $ \theta = 2 \arctan K$ and the controlled part of the gate becomes
\be
U_{\rm CZS} (\theta, \phi, 0) =
\begin{bmatrix}
1 & 0 & 0 & 0 \\
0 & \cos \theta & e^{-i\phi} \sin \theta & 0 \\
0 & e^{i\phi} \sin \theta & - \cos \theta & 0 \\
0 & 0 & 0 & - 1
\end{bmatrix}.
\label{eq:UCZS_gamma0}
\ee
The gate time becomes $t_{\rm gate} = \pi / \sqrt{1 + K^2} \lambda$. This is faster than the corresponding individual CZ gates, which would take $t_{\rm gate} = \pi / \lambda$ and $t_{\rm gate} = \pi / K \lambda$, respectively, on their own.


\subsubsection{Time-dependent parameters}

In the derivation of Eqs.~(\ref{eq:CCZS})-(\ref{eq:parameter}), which constitute the main result of this section, we assumed for simplicity that the coupling strengths $\lambda_1$, $\lambda_2$ and the detuning $\delta$ were constants. However, in actual experiments, at least the coupling strengths $\lambda_1$ and $\lambda_2$ will need to be turned on and off, and this will not be done with a perfect step function. Fortunately, it is feasible to vary these parameters in time as long as they have the same time dependence. The principle is the same as for nonadiabatic holonomic gates~\cite{Sjoqvist2012, Abdumalikov2013, Danilin2018, Egger2019}. As \eqref{eq:2level} shows, we can thus construct arbitrary rotations in the space spanned by the states $\ket{101}$ and $\ket{110}$, where the angles $\theta$, $\phi$ are controlled by the relative strengths of the two CZ gates being applied simultaneously [see \eqref{eq:ThetaPhiRelation}].


\subsection{Decomposition into two-qubit gates}
\label{sec:DecompositionCCZS}

The three-qubit CCZS gate entangles all three qubits. It can thus not be written as the simultaneous application of a two-qubit gate to two of the qubits and a single-qubit gate to the third qubit. Instead, decomposing CCZS into single- and two-qubit gates requires the consecutive application of several such gates. Inspired by decompositions for the quantum-optical Fredkin gate~\cite{Milburn1989, Nielsen2000}, we find that CCZS can be realized through the consecutive application of three two-qubit gates:
\be
{\rm CCZS} (\theta, \phi, \gamma) = \mathrm{XY}_{12} (\theta, \frac{\pi}{2} - \phi) \cdot \mathrm{CZ}_{01} (\gamma) \cdot \mathrm{XY}^\dag_{12} (\theta, \frac{\pi}{2} - \phi),
\label{eq:CCZSDecomposition}
\ee
where
\be
\mathrm{XY}(\theta, \phi) = 
\begin{bmatrix}
1 & 0 & 0 & 0 \\
0 & \cos \frac{\theta}{2} & i \sin \frac{\theta}{2} e^{i \phi} & 0 \\
0 & i \sin \frac{\theta}{2} e^{- i \phi} & \cos \frac{\theta}{2} & 0 \\
0 & 0 & 0 & 1
\end{bmatrix}
\ee
and
\be
\mathrm{CZ}(\gamma) = 
\begin{bmatrix}
1 & 0 & 0 & 0 \\
0 & 1 & 0 & 0 \\
0 & 0 & 1 & 0 \\
0 & 0 & 0 & - e^{i \gamma}
\end{bmatrix}.
\ee
Here, the two-qubit $\mathrm{XY}(\theta, \phi)$ gate~\cite{Ganzhorn2019, Abrams2020} is generated by an exchange-type interaction, e.g., $\mathrm{XY} (\pi, 0) = \mathrm{iSWAP}$.

\begin{figure}[]
\includegraphics[width=\linewidth]{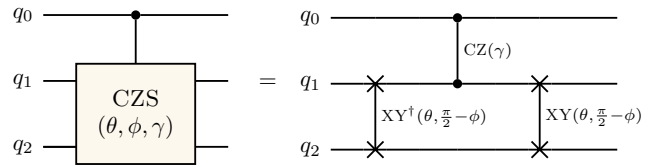}
\caption{Decomposition of the three-qubit CCZS gate into two-qubit gates. Note that for the case of a linear chain [see \figpanel{fig:PrincipleCCZS}{a}], the control qubit $q_0$ in the CCZS gate is the middle qubit, but the decomposition into two-qubit gates requires $q_1$ to be the middle qubit, since it has to interact with both $q_0$ and $q_2$.
\label{fig:DecompositionCCZS}}
\end{figure}

The decomposition in \eqref{eq:CCZSDecomposition} is illustrated in \figref{fig:DecompositionCCZS}. From that illustration, it becomes clear that this decomposition requires re-labelling the qubits in a linear chain to work. For the case of a linear chain, the control qubit in the CCZS gate is the middle qubit, but in the decomposition given here, the control qubit must be one of the outer qubits.


\subsection{Constructing other three-qubit gates}
\label{sec:CCZSOther3qGates}

We now check how the CCZS gate family is related to some well-known three-qubit gates: the Fredkin (controlled-SWAP)~\cite{Fredkin1982, Milburn1989, Smolin1996, Nielsen2000}, iFredkin (controlled-iSWAP)~\cite{Dallaire-Demers2016, Liebermann2017, Rasmussen2020a}, and Toffoli (controlled-controlled-NOT)~\cite{Toffoli1980, Nielsen2000} gates. Just like the CCZS gate, these three-qubit gates can be written on the form of \eqref{eq:CCZS}, but with $U_{\rm CZS} (\theta, \phi, \gamma)$ replaced by other controlled two-qubit unitary operations:
\bea
U_{\rm Fredkin} =
\begin{bmatrix}
1 & 0 & 0 & 0 \\
0 & 0 & 1 & 0 \\
0 & 1 & 0 & 0 \\
0 & 0 & 0 & 1
\end{bmatrix}, 
\label{eq:UFredkin}
\\
U_{\rm iFredkin} =
\begin{bmatrix}
1 & 0 & 0 & 0 \\
0 & 0 & i & 0 \\
0 & i & 0 & 0 \\
0 & 0 & 0 & 1
\end{bmatrix}, 
\label{eq:UiFredkin}
\\
U_{\rm Toffoli} =
\begin{bmatrix}
1 & 0 & 0 & 0 \\
0 & 1 & 0 & 0 \\
0 & 0 & 0 & 1 \\
0 & 0 & 1 & 0
\end{bmatrix}.
\label{eq:UToffoli}
\eea

Comparing with $U_{\rm CZS} (\theta, \phi, \gamma)$ in \eqref{eq:UCZS}, it is clear that it never coincides with $U_{\rm Toffoli}$ in \eqref{eq:UToffoli}, since the two off-diagonal elements in the lower right corner of $U_{\rm CZS} (\theta, \phi, \gamma)$ are zero for all values of $\theta$, $\phi$, and $\gamma$. Noting that a Toffoli gate is formed by sandwiching a controlled-controlled-Z (CCZ) gate~\cite{Fedorov2012} between two Hadamard gates on qubit 2 does not help. The controlled unitary of the CCZ gate is
\be
U_{\rm CZ} = 
\begin{bmatrix}
1 & 0 & 0 & 0 \\
0 & 1 & 0 & 0 \\
0 & 0 & 1 & 0 \\
0 & 0 & 0 & - 1
\end{bmatrix}.
\label{eq:UCCZ}
\ee
To obtain the $-1$ in this matrix from $U_{\rm CZS} (\theta, \phi, \gamma)$ requires $\gamma = 0$, which is easy, but we see from \eqref{eq:UCZS_gamma0} that the two middle diagonal elements in $U_{\rm CZS} (\theta, \phi, \gamma)$ then always will have opposite signs, which does not match the CCZ gate. Indeed, for $\theta = 0$, we simply have the two-qubit CZ gate acting on qubits 0 and 1.

\begin{figure}[]
\includegraphics[width=\linewidth]{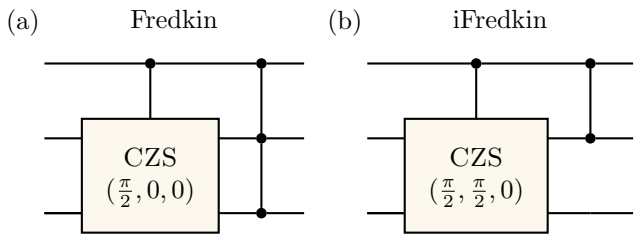}
\caption{Converting CCZS gates into other three-qubit gates. (a) Constructing a Fredkin gate with a $\mathrm{CCZS} (\pi/2, 0, 0)$ gate and a CCZ gate. (b) Constructing an iFredkin gate from a $\mathrm{CCZS} (\pi/2, \pi/2, 0)$ gate and a CZ gate. 
\label{fig:FredkinGatesWithCCZS}}
\end{figure}

The Fredkin gate in \eqref{eq:UFredkin} also cannot be directly implemented by the CCZS gate. For the $1$ in the lower right corner of the $U_{\rm Fredkin}$ to match $U_{\rm CZS} (\theta, \phi, \gamma)$, $\gamma = \pm \pi$ is necessary, but then all off-diagonal elements in $U_{\rm CZS} (\theta, \phi, \gamma)$ become zero. However, we note that the Fredkin gate can be constructed by combining a CCZ gate and $\mathrm{CCZS} (\pi/2, 0, 0)$ (i.e., $\lambda_1 = - \lambda_2$ and $\delta = 0$), as shown in \figpanel{fig:FredkinGatesWithCCZS}{a}. Since an implementation of the Fredkin gate using only single- and two-qubit gates requires at least five two-qubit gates~\cite{Smolin1996}, and a CCZ gate can be implemented using three two-qubit gates~\cite{Fedorov2012}, the construction with the CCZS gate (which on its own is at least as fast as a two-qubit gate) constitutes an improvement.

Using exactly the same reasoning as for the Fredkin gate in the preceding paragraph, we see that the iFredkin gate in \eqref{eq:UiFredkin} also cannot be directly implemented by the CCZS gate. However, it is sufficient to add a single two-qubit CZ gate after $\mathrm{CCZS} (\pi/2, \pi/2, 0)$ to fix this, as shown in \figpanel{fig:FredkinGatesWithCCZS}{b}. Since implementing the iFredkin gate using only single- and two-qubit gates requires at least four two-qubit gates, while the CCZS gate is at least as fast as a two-qubit gate, our construction at least halves the time required for the iFredkin gate, which is a natural operation in, e.g., simulations of the Fermi-Hubbard model~\cite{Dallaire-Demers2016}.


\subsection{Rapid creation of large entangled states}
\label{sec:CCZSEntanglement}

Having demonstrated that the CCZS gate is fast and that it entangles three qubits, we now show how this gate and its generalization to more than three qubits can be applied to rapidly generate some particular large entangled states. The ability to create entanglement~\cite{Horodecki2009} is crucial for both quantum information processing~\cite{Nielsen2000, Briegel2009, Preskill2012} and quantum communication~\cite{Gisin2007}. Lately, the creation of entanglement between (several) tens of qubits has been used to demonstrate the capabilities of quantum processors on multiple platforms: superconducting circuits~\cite{Wang2018, Mooney2019, Wei2020, Mooney2021, Mooney2021a, Yang2021}, photonic systems~\cite{Wang2018}, ion traps~\cite{Friis2018, Pogorelov2021}, and neutral atoms~\cite{Omran2019}. 


\subsubsection{Greenberger-Horne-Zeilinger states}

\begin{figure}[]
\includegraphics[width=\linewidth]{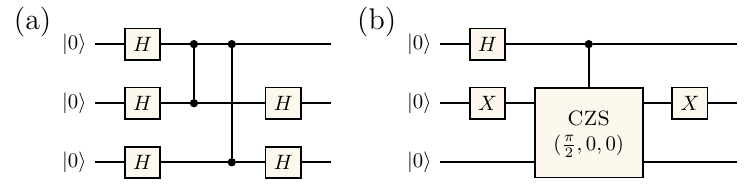}
\caption{Quantum circuits for generating the three-qubit GHZ state $\mleft( \ket{000} + \ket{111} \mright) / \sqrt{2}$. (a) A circuit using single-qubit gates and two-qubit CZ gates. (b) A circuit using single-qubit gates and one CCZS gate.
\label{fig:GHZ}}
\end{figure}

Greenberger-Horne-Zeilinger (GHZ) states~\cite{Greenberger1989, Bouwmeester1999} are entangled states of $N$ qubits on the form
\be
\ket{\Psi_{\rm GHZ}} = \frac{1}{\sqrt{2}} \mleft( \ket{0}^{\otimes N} + \ket{1}^{\otimes N} \mright).
\ee
To generate a GHZ state with $N = 3$ qubits using only single- and two-qubit gates requires at least two two-qubit gates, e.g., two CNOT gates, two iSWAP gates~\cite{Neeley2010}, or two CZ gates [see \figpanel{fig:GHZ}{a}]. However, this state can also be generated using only a single CCZS gate and a few single-qubit gates, using the circuit shown in \figpanel{fig:GHZ}{b}. Starting from $\ket{000}$, applying single-qubit Hadamard and X gates to the first and second qubit, respectively, creates the state
\be
\ket{\psi} = \frac{1}{\sqrt{2}} \mleft( \ket{0} + \ket{1} \mright) \ket{1}\ket{0}. 
\ee
Then, applying $\mathrm{CCZS} (\theta = \pi / 2, \phi = 0, \gamma = 0)$, which is achieved for $\lambda_1 = - \lambda_2$ and $\delta = 0$, results in a controlled SWAP of the second and third qubit [see \eqref{eq:UCZS_phi}], yielding the state
\be
\mathrm{CCZS} (\pi / 2, 0, 0) \ket{\psi} = \frac{1}{\sqrt{2}} \mleft( \ket{010} + \ket{101} \mright),
\ee
which is transformed to the three-qubit GHZ state by applying an X gate on the second qubit. We remark that the phase acquired by the doubly excited state of the second and third qubits does not affect the state $\ket{\psi}$ that we apply the CCZS gate to.

Since the $\mathrm{CCZS} (\pi / 2, 0, 0)$ is $\sqrt{2}$ faster than the CZ gates that can be implemented with the interactions from which it is constructed (see \secref{sec:Examples3QubitGatesCCZS}), the circuit with the CCZS gate in \figpanel{fig:GHZ}{b} generates the three-qubit GHZ state $2 \sqrt{2}$ times faster than the circuit with the two CZ gates in \figpanel{fig:GHZ}{a}, provided that the time required for single-qubit gates is negligible. If one instead uses the circuit with two iSWAP gates~\cite{Neeley2010}, the circuit with the CCZS gate is twice as fast, since the iSWAP is $\sqrt{2}$ faster than a CZ gate in the setups we consider here (the coupling between $\ket{11}$ and $\ket{02}$ is $\sqrt{2}$ stronger than the coupling between $\ket{01}$ and $\ket{10}$ for weakly anharmonic qubits, but the iSWAP only require half the oscillation between states that CZ does; see, e.g., Ref.~\cite{Ganzhorn2020}).


\subsubsection{Dicke states}
\label{sec:DickeCCZS}

Another class of entangled states is the W states
\be
\ket{\Psi_{\rm W}} = \frac{1}{\sqrt{N}} \mleft( \ket{100 \ldots 0} + \ket{010 \ldots 0} + \ldots + \ket{000 \ldots 1} \mright),
\ee
which cannot be converted into GHZ states by local operations and classical communication~\cite{Dur2000}. The W states are in turn a subset $\ket{D_N^1}$ of the symmetric Dicke states $\ket{D_N^k}$~\cite{Dicke1954, Haroche2006, Shammah2018}, which are equally weighted superpositions of all permutations of $N$-qubit states with $k$ excitations:
\be
\ket{D_N^k} = \frac{1}{\sqrt{N \choose k}} \mathcal{S} \mleft[ \ket{0}^{\otimes (N - k)} \otimes \ket{1}^{\otimes k} \mright],
\label{eq:DickeDefinition}
\ee
with $\mathcal{S}$ the symmetrization operator.

Dicke states have important applications in quantum metrology~\cite{Pezze2018, Paulisch2019} and quantum networks~\cite{Prevedel2009, Wieczorek2009, Chiuri2012, Miguel-Ramiro2020}. Recently, it has also been shown that, for combinatorial optimization problems, symmetric Dicke states representing a superposition of all feasible solutions can give advantages when used as the initial state in the quantum alternating operator ansatz~\cite{Cook2020, Bartschi2020}.

The Dicke states arise naturally when $N$ identical atoms are collectively coupled to a harmonic mode~\cite{Haroche2006, Hume2009}. However, since the photon or phonon number of the harmonic mode is difficult to control, alternative protocols for Dicke-state generation have been proposed~\cite{Stockton2004, Xiao2007, Lamata2013, Ivanov2013, Wu2017}. For deterministic preparation of a symmetric Dicke state on a quantum computer, using a sequence of single- and two-qubit gates, it has been shown that constructing $\ket{D_N^k}$ requires a quantum circuit with depth $\mathcal{O}(N)$ containing at least $\mathcal{O}(kN)$ gates~\cite{Chakraborty2014, Bartschi2019, Mukherjee2020}.

In this subsection, we show how to rapidly create large symmetric Dicke states by generalizing the interaction underpinning the CCZS gate to more qubits. As a concrete example, we show that we can create the state $\ket{D_5^3}$ using only two rounds of simultaneous CZ gates and two single-qubit operations, while the quantum circuit in Ref.~\cite{Bartschi2019} for creating the same state includes five three-qubit gates and 21 two-qubit gates applied sequentially.


\paragraph{Hamiltonian and dynamics}
\label{sec:DickeHamiltonianAndDynamics}

We consider a system composed of a central qubit $0$ and its $N$ nearest neighbours $\{ j \}$ (the three-qubit system in \figref{fig:PrincipleCCZS} had $\{ j \} = \{1, 2\}$). For each qubit $i$, we take into account the three lowest energy levels $\ket{0_i}$, $\ket{1_i}$, and $\ket{2_i}$, with energies 0, $\omega_i$, and $2 \omega_i + \alpha_i$, respectively, where $\alpha_i$ is the anharmonicity. In this system, a CZ gate between qubit $0$ and one of its neighbours $j$ can be applied by activating the $\ket{2_0 0_j} \leftrightarrow \ket{1_0 1_j}$ transition. Assuming these transitions are resonant, the system Hamiltonian with the transitions switched on is, in the interaction picture,
\be
H = \sigma_0^{21} \sum_{j = 1}^N \lambda_j \sigma_j^{01} + \text{H.c.},
\label{eq:HNCZ}
\ee
where $\sigma_i^{nm} = \ketbra{n}{m}_i$ and $\lambda_j$ is the interaction strength for the $\ket{2_0 0_j} \leftrightarrow \ket{1_0 1_j}$ transition.

For simplicity, we assume all interactions equally strong: $\lambda_j \equiv \lambda$. We can then introduce the collective spin operator $J^- = \sum_{j = 1}^N \sigma_j^{01}$ and rewrite \eqref{eq:HNCZ} as
\be
H = \lambda \sigma_0^{21} J^- + \text{H.c.}, 
\label{eq:Tavis}
\ee
which is reminiscent of the Tavis--Cummings (Dicke) model~\cite{Tavis1968, Klimov2009}, where a harmonic oscillator couples
to $N$ identical atoms. Due to the anharmonicity of qubit $0$, the models are equivalent in the limit where the harmonic oscillator only hosts a single photon.

The neighbouring qubits $1$ to $N$ are symmetric under permutation and can thus be described by the Dicke states in \eqref{eq:DickeDefinition}. In this basis, the matrix elements of the operator $J^+ = \mleft( J^- \mright)^\dag$ are~\cite{Noguchi2012}
\bea
\brakket{D_N^{k+1}}{J^+}{D_N^k} &=& \mleft( N - k \mright) \sqrt{\frac{{N \choose k}}{{N \choose k + 1}}} \nn\\
&=& \sqrt{\mleft( N - k \mright) \mleft( k + 1 \mright)} \equiv G_N^k.
\label{eq:DickeMatrixElements}
\eea
We can thus interpret the second term in \eqref{eq:Tavis} as qubit $0$ being de-excited from $\ket{2_0}$ to $\ket{1_0}$ while the state of the neighbouring qubits changes from $\ket{D_N^k}$ to $\ket{D_N^{k+1}}$. The total excitation number $k+2$ is conserved.

Since subspaces with different numbers of excitations are decoupled, we can limit ourselves to the subspace with $k+2$ excitations, which is spanned by the basis $\ket{1_0} \ket{D_N^{k+1}}$, $\ket{2_0} \ket{D_N^k}$. Using Eqs.~(\ref{eq:Tavis})-(\ref{eq:DickeMatrixElements}), we see that the dynamics in this subspace is generated by
\be
\exp \mleft(- i t \lambda G_N^k \tilde{\sigma}_x \mright) = \cos \mleft( t \lambda G_N^k \mright) - i \sin \mleft( t \lambda G_N^k \mright) \tilde{\sigma}_{x},
\label{eq:DickeDynamics}
\ee
where $\tilde{\sigma}_{x}$ is defined in the basis of the subspace.


\paragraph{Generalizing the CCZS gate to more than three qubits}

We can now understand how the CCZS gate generalizes to more qubits. For $N=2$ neighbouring qubits, starting in the computational subspace, i.e., with qubit $0$ in state $\ket{1_0}$, we can always find a time $t$ when the occupation of $\ket{2_0}$ is zero, since $G_2^1 = G_2^0 = \sqrt{2}$. This is the case analyzed in the preceding subsections. For  $N > 2$, the coefficients $G_N^k$ are not equal or commensurate, making it impossible to confine the central qubit to the computational subspace and create an $(N+1)$-qubit gate according to the same principle as the CCZS gate. What could be done is to apply a single-qutrit operation on qubit $0$ that takes $\ket{0_0}$ to $\ket{1_0}$ and $\ket{1_0}$ to $\ket{2_0}$, let the system evolve for some time $t$ according to $\eqref{eq:DickeDynamics}$, and then apply the inverse of the single-qutrit operation to qubit $0$ to bring it back to the computational subspace. While this would be an $(N+1)$-qubit gate, it appears too complicated to find immediate applications.


\paragraph{Creating a five-qubit Dicke state}
\label{sec:5QubitDicke}

Instead of constructing a general $(N+1)$-qubit gate, we therefore focus on preparing symmetric Dicke states by starting in a specific subspace. As an illustrative example, we consider the case $N = 4$ and the target state
\be
\ket{D_5^3} = \sqrt{\frac{3}{5}} \ket{1_0} \ket{D_4^2} + \sqrt{\frac{2}{5}} \ket{0_0} \ket{D_4^3}, 
\label{eq:D53}
\ee
which also was used as an example in Ref.~\cite{Bartschi2019}.

\begin{figure}
\includegraphics[width=\linewidth]{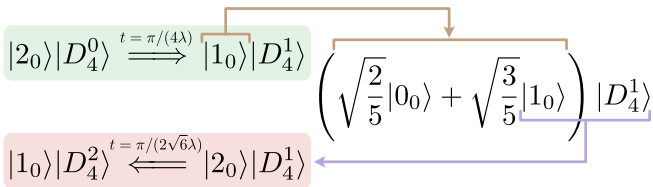}
\caption{The steps for preparing the state $\sqrt{\frac{3}{5}} \ket{1_0} \ket{D_4^2} + \sqrt{\frac{2}{5}} \ket{0_0} \ket{D_4^1}$ using the generalization of the CCZS gate to a 5-qubit system.
\label{fig:DickePreparation}}
\end{figure}

First, we note that since $\ket{D_4^3}$ can be created by applying X gates to all qubits in the state $\ket{D_4^1}$, and the state $\ket{D_4^2}$ is unchanged by those gates, the problem reduces to preparing the state $\sqrt{\frac{3}{5}} \ket{1_0} \ket{D_4^2} + \sqrt{\frac{2}{5}} \ket{0_0} \ket{D_4^1}$. The procedure for doing so is illustrated in \figref{fig:DickePreparation}. We first explain how to obtain $\ket{D_4^2}$:
\begin{enumerate}
\item
First, we prepare the initial state $\ket{20000} = \ket{2_0} \ket{D_4^0}$ by single-qutrit operations on qubit 0. This puts the system in the subspace $k = 0$ spanned by $\ket{1_0} \ket{D_4^1}$ and $\ket{2_0} \ket{D_4^0}$. Turning on the interaction and letting the system evolve for a time $t = \pi / (4 \lambda)$, four times faster than a two-qubit CZ gate, we arrive at $\ket{1_0} \ket{D_4^1}$. We remark that if we have the system tuned to have different interaction strengths $\lambda_j$ like in tripod systems~\cite{Unanyan1999, Mousolou2018}, we can create arbitrarily weighted superpositions of the $N$-qubit states with one excitation instead of the symmetric superposition that is the symmetric Dicke state.
\item
Next, we flip qubit $0$ to $\ket{2_0}$ such that the system state becomes $\ket{2_0} \ket{D_4^1}$. This puts the system in the $k = 1$ subspace spanned by $\ket{1_0} \ket{D_4^2}$ and $\ket{2_0} \ket{D_4^1}$. Turning on the interaction again for a time $t = \pi / (2 \sqrt{6} \lambda)$, we arrive at $\ket{1_0} \ket{D_4^2}$.
\end{enumerate}

To create the superposition state in \eqref{eq:D53}, we carry out step 1 as above. Then we rotate qubit $0$ to the superposition state $\sqrt{\frac{2}{5}} \ket{0_0} + \sqrt{\frac{3}{5}} \ket{1_0}$ and flip $\ket{1_0}$ to $\ket{2_0}$, yielding the system state $\mleft( \sqrt{\frac{2}{5}} \ket{0_0} + \sqrt{\frac{3}{5}} \ket{2_0} \mright) \ket{D_4^1}$.  Turning on the interaction as in step 2 above, the part of the superposition containing $\ket{0_0}$ is decoupled from the dynamics, while the part containing $\ket{2_0}$ reaches $\ket{1_0} \ket{D_4^2}$ as before. Finally, applying X gates to the four neighbouring qubits yields the state in \eqref{eq:D53}.

In total, our scheme requires seven single-qubit operations (four of them simultaneous) and two applications of the interaction that yields CZ gates. The total time spent on these CZ interactions is less than that of a single two-qubit CZ gates. This fast creation of the entangled state in \eqref{eq:D53} should be contrasted with the quantum circuit for the same task given in Ref.~\cite{Bartschi2019}, which contained five three-qubit gates and 21 two-qubit gates, applied sequentially.


\paragraph{Rapid creation of a large W state}

\begin{figure}
\includegraphics[width=0.9\linewidth]{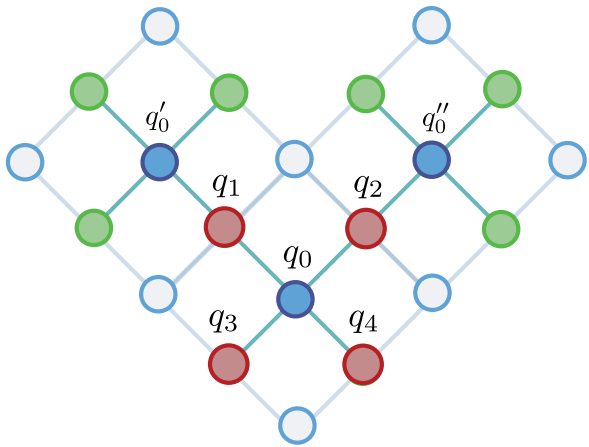}
\caption{Creating large W states rapidly on a square grid of qubits. We first prepare qubit $0$ ($q_0$) in its second excited state and carry out the rest of step 1 in \secref{sec:5QubitDicke} to create $\ket{D_4^1}$ on the neighbouring qubits (red circles). Next, we swap each of the neighbouring qubits with the qubit next to them that is farthest from the centre qubit (filled blue circles except $q_0$). Flipping the $\ket{1}$ part of the state for these new centre qubits to $\ket{2}$ and having them interact with their nearest neighbours creates $\ket{D_4^1}$ on those neighbours (green and red circles). For brevity, the new cells starting from $q_3$ and $q_4$ are not shown. The result after these steps (two rounds of single or simultaneous single-qubit operations, three rounds of simultaneous two-qubit gates) is the 16-qubit W state $\ket{D_{16}^1}$.
\label{fig:WStatesCZScalingUp}}
\end{figure}

The size of the Dicke state that can be efficiently prepared is determined by the number of neighbouring qubits to which the central qubit is coupled. In general, scaling up arbitrary Dicke states is hard~\cite{Kobayashi2014}. However, in our scheme, large W states are easy to construct even with limited connectivity, e.g., in a square grid of qubits, as shown in \figref{fig:WStatesCZScalingUp}. The method outlined there is straightforward to adapt for other connectivities.


\subsection{Experimental feasibility}
\label{sec:ExpCCZS}

To determine how well the CCZS gate is likely to work in actual experiments, we now turn to simulating two specific possible experimental implementations of the gate. We consider two gate schemes commonly used to perform CZ gates in superconducting circuits by turning on and off an interaction between the two-qubit states $\ket{11}$ and $\ket{02}$ or $\ket{20}$. Similar schemes used for CZ gates on other quantum-computing platforms should be equally feasible for realizing the CCZS gate.

In the first scheme, the two outer qubits in the three-qubit chain are tunable~\cite{DiCarlo2009, Barends2019, Andersen2019, Negirneac2021}. To activate the gate, they are tuned such that the $\ket{110}$ and $\ket{101}$ states both become resonant with the $\ket{200}$ state. In some implementations, tunable couplers between the qubits are also adjusted to further control the coupling~\cite{Foxen2020, Sung2021, Xu2021}. In all these cases, the interaction strengths $\lambda_1$ and $\lambda_2$ will be limited to being in phase, i.e., $\phi = \pi$ [see \eqref{eq:ThetaPhiRelation}]. In the cases without tunable couplers, the parameter $\theta$ [see \eqref{eq:ThetaPhiRelation}] is fixed by the coupling strengths in the hardware and cannot be tuned in situ.

In the second scheme, the neighbouring qubits in the chain are connected via a tunable coupler, which itself is a qubit~\cite{McKay2016, Roth2017, Bengtsson2020, Ganzhorn2020}. To activate a CZ gate, the coupler, which is detuned from the qubits it is connected to, is parametrically modulated with a modulation frequency close to the difference in frequency between the states $\ket{11}$ and $\ket{02}$ or $\ket{20}$. In this case, the interaction strengths $\lambda_1$ and $\lambda_2$ are determined by the phase and amplitude of the modulation of the coupler; they can thus be tuned over a wide range to implement different parameters for the CCZS gate. We note that a CZ gate also can be implemented in a similar fashion between a fixed-frequency qubit and a parametrically modulated qubit~\cite{Reagor2018, Caldwell2018}, but we do not simulate that case here.

To characterize the performance of the gates, we calculate the average gate fidelity~\cite{Nielsen2002, Willsch2020}
\bea
F &=& \int \id \ket{\psi} \brakket{\psi}{U^\dag M \ketbra{\psi}{\psi} M^\dag U}{\psi} \\
&=& \frac{\abssq{\tr{M U^\dag}} + \tr{M^\dag M}}{n(n+1)},
\eea
where $U$ is the ideal gate operation that we wish to implement, $M$ is the gate operation that we actually implement, and $n$ is the dimension of the computational subspace ($n = 2^2 = 4$ for the CZ gates and $n = 2^3 = 8$ for the CCZS gate). The aim of the numerical simulations below is to show that high-fidelity three-qubit CCZS gates can be obtained in a straightforward way, without optimizing pulse shapes, etc., compared to the constituent two-qubit gates.


\subsubsection{Tunable qubits}
\label{sec:ExpCCZSTunableQubits}

\begin{figure}[]
\includegraphics[width=\linewidth]{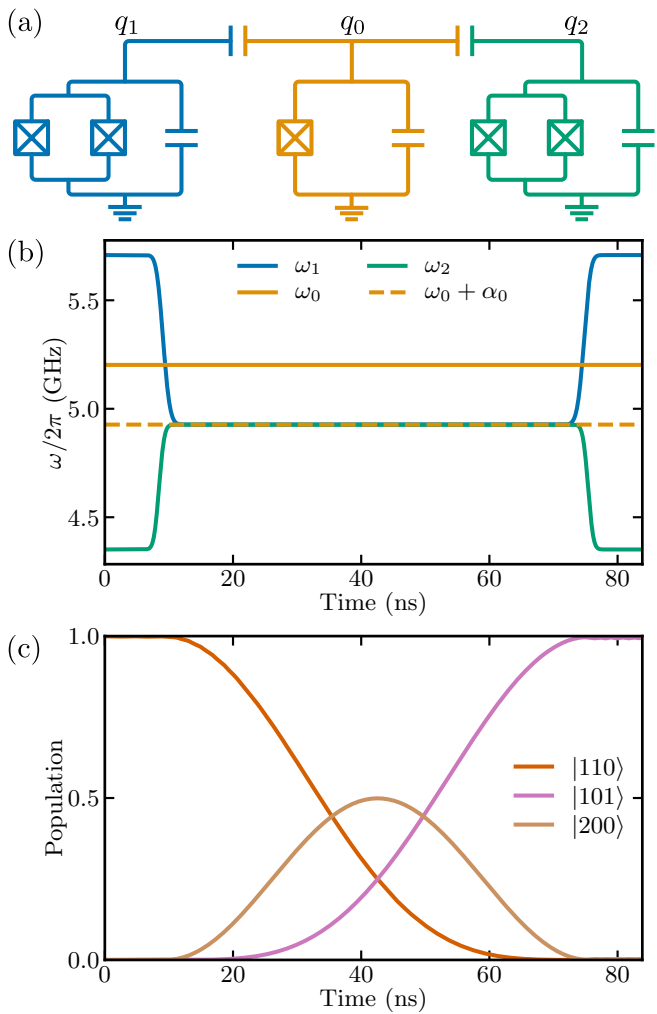}
\caption{Implementing a CCZS gate with tunable qubits.
(a) A sketch of the setup, with qubit numbering following the convention established in \figpanel{fig:PrincipleCCZS}{a}. The superconducting qubits are transmon qubits~\cite{Koch2007}, which are nonlinear $LC$ oscillators where the nonlinear inductances are provided by Josephson junctions (boxes with crosses in the sketch). When two Josephson junctions are combined in a loop [a superconducting quantum interference device (SQUID)], the effective inductance, and thus the qubit frequency, can be tuned by controlling the magnetic flux through the loop.
(b) The tuning of the qubit energies used to implement a high-fidelity $\mathrm{CCZS} (\pi / 2, \pi, 0)$ gate.
(c) Population of the states $\ket{110}$, $\ket{200}$, and $\ket{101}$ during the $\mathrm{CCZS} (\pi / 2, \pi, 0)$ gate when the initial state is $\ket{110}$. As \eqref{eq:UCZSpi2pi0} shows, the main effect of this gate is to swap the population between qubits 1 and 2 if qubit 0 is in its excited state.
\label{fig:TunableQubitsCCZS}}
\end{figure}

We first consider the setup with two tunable qubits 1 and 2 on each side of the fixed-frequency qubit 0 as shown in \figpanel{fig:TunableQubitsCCZS}{a}. To test the performance under realistic conditions, we use parameters close to the experiment in Ref.~\cite{Andersen2019}. Before the gate is turned on, the qubit energies are
\be
\mleft[ \omega_0, \omega_1, \omega_2 \mright] = 2 \pi \times \mleft[ 5.202, 5.708, 4.350 \mright]\,\text{GHz}.
\label{eq:OmegaValuesCCZSTunableQubits}
\ee
For qubits 0 and 1, these are also their maximum energies; for qubit 2, the maximum energy is set to $2 \pi \times \unit[4.927]{GHz}$. The qubit anharmonicities are
\be
\mleft[ \alpha_0, \alpha_1, \alpha_2 \mright] = - 2 \pi \times \mleft[ 275.2, 261.1, 277.3 \mright]\,\text{MHz}
\ee
and the couplings between the qubits are
\be
\lambda_1 = \lambda_2 = \unit[2 \pi \times \sqrt{2} \times 3.8]{MHz}.
\label{eq:LambdaValuesCCZSTunableQubits}
\ee
The factor $\sqrt{2}$ appears in \eqref{eq:LambdaValuesCCZSTunableQubits} since we use $\lambda_1$ and $\lambda_2$ to denote the coupling strengths for $\ket{11} \leftrightarrow \ket{20}$ transitions instead of the coupling strengths for $\ket{10} \leftrightarrow \ket{01}$ transitions, which are the parameters that are actually given as input to the simulation.

To activate the gate, the energies $\omega_1$ and $\omega_2$ of qubits 1 and 2 are tuned into resonance with $\omega_0 + \alpha_0$ as shown in \figpanel{fig:TunableQubitsCCZS}{b}. To account for the finite response time of the drive line, the pulse used for tuning the qubit energies is the convolution of a rectangular pulse of length $t_{\rm gate}$ (the gate time) and a Gaussian pulse centered in the middle of the rectangular pulse with standard deviation $\sigma = \unit[1]{ns}$.

We first check that we can tune up CZ gates between qubits 1 or 2 and qubit 0 by tuning just one of qubits 1 and 2 to the relevant resonance. We find that we can achieve $F > \unit[99.99]{\%}$ for the CZ gate between qubits 0 and 2 with a gate time $t_{\rm gate, CZ_{02}} = \unit[93.0]{ns}$ and, similarly, $F = \unit[99.82]{\%}$ for the CZ gate between qubits 0 and 1 with a gate time $t_{\rm gate, CZ_{01}} = \unit[94.0]{ns}$.

We then tune up the $\mathrm{CCZS} (\pi / 2, \pi, 0)$ gate [see \eqref{eq:UCZSpi2pi0}] by reducing the gate time and synchronizing the tuning of both qubits 1 and 2 to the resonance for the gate. Note that $\theta = \pi / 2$ is set by the fixed coupling strengths $\lambda_1 = \lambda_2$ and cannot be changed. The best gate fidelity we find is $F = \unit[99.42]{\%}$ for the gate time $t_{\rm gate, CCZS} = \unit[66.8]{ns} \approx t_{\rm gate, CZ} / \sqrt{2}$ [see \figpanels{fig:TunableQubitsCCZS}{b}{c}]. Note that the fidelity values for both the CCZS and CZ gates here are calculated without including any effects of decoherence.

We attribute the deviation from \unit[100]{\%} gate fidelity for the three-qubit gate to a combination of factors. One is imperfections arising when tuning qubits 1 and 2 in and out of the resonance. During that time, qubit 1 crosses the frequency of qubit 0, which may cause leakage by briefly activating the $\ket{01} \leftrightarrow \ket{10}$ transition for these qubits instead of the desired $\ket{11} \leftrightarrow \ket{20}$. We note that qubit 2, which is below qubit 0 in frequency, does not have the same potential problem; this may explain why the $\text{CZ}_{02}$ gate has higher fidelity than the $\text{CZ}_{01}$ gate. Furthermore, tuning qubits 1 and 2 from different frequencies into the resonance appears to affect the parameter $\phi$, making it deviate from $\pi$ and thus lowering the gate fidelity. To improve the tuning of the qubit energies, one could try methods developed for nonadiabatic holonomic gates~\cite{Liu2019}.

An additional source of error may be that the states $\ket{001}$, $\ket{100}$, and $\ket{010}$ form a $\Lambda$ system during the gate operation, with $\ket{001}$ and $\ket{010}$ having the same energy. Although $\ket{100}$ is detuned from the other two states by the anharmonicity $\alpha_0$, there will still be a small effective coupling between $\ket{001}$ and $\ket{010}$ that can contribute to lowering the gate fidelity. This effect can be reduced by increasing the anharmonicity.

\begin{figure*}[]
\includegraphics[width=\linewidth]{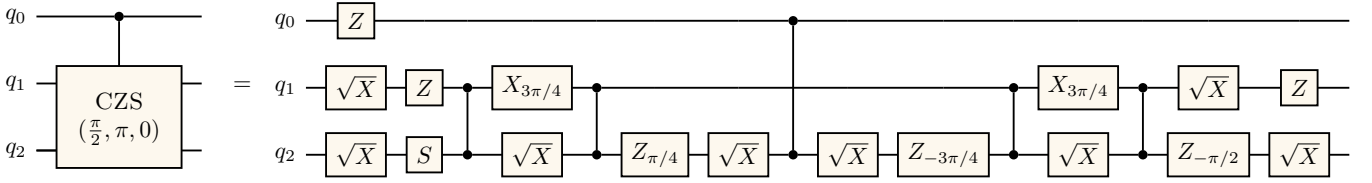}
\caption{Decomposition of the $\mathrm{CCZS} (\pi / 2, \pi, 0)$ gate into single-qubit and CZ gates, obtained using Qiskit~\cite{Qiskit}. For the single-qubit gates, we use the notation $S = \sqrt{Z} = Z_{\pi/2}$ and $\sqrt{X} = X_{\pi/2}$. Note that this decomposition requires qubit 2 being placed in the middle of the linear chain, since it has to perform CZ gates with both qubit 0 and qubit 1.
\label{fig:DecompositionCCZSPi2Pi0}}
\end{figure*}

To further put the time gained by performing the three-qubit gate in perspective, we show in \figref{fig:DecompositionCCZSPi2Pi0} the decomposition of the $\mathrm{CCZS} (\pi / 2, \pi, 0)$ gate into single-qubit gates and two-qubit CZ gates between the middle qubit and its neighbours. Note that this is different from the decomposition in \figref{fig:DecompositionCCZS}, which assumes access to a parameterized XY gate in addition to the CZ gates we have here. From the decomposition in \figref{fig:DecompositionCCZSPi2Pi0}, we see that five sequential CZ gates would be needed to implement this three-qubit gate in the setup at hand. Even if we assume that single-qubit gates take negligible time compared to two-qubit gates, this still means that we gain more than a factor 7 in gate time by implementing the three-qubit gate using our scheme.


\subsubsection{Tunable couplers}
\label{sec:ExpCCZSTunableCouplers}


\paragraph{Setup and operation}

\begin{figure}[]
\includegraphics[width=\linewidth]{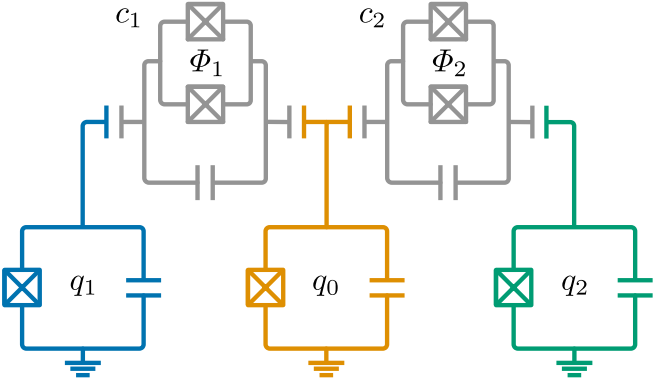}
\caption{Setup for implementing the CCZS gate with tunable couplers $c_1$ and $c_2$ connecting the qubits. The frequency of each coupler $j$ is tuned by changing the magnetic flux $\Phi_j$ through the loop formed by the two Josephson junctions.
\label{fig:CCZSTunableCouplerSetup}}
\end{figure}

\begin{table}
\centering
\caption{Parameter values used in \eqref{eq:HTunableCoupler} for the simulations of implementing CCZS gates with tunable couplers. Here, $\omega_{c_j}^0$ is the maximum frequency of coupler $j$. All values are given in units of GHz.
\label{tab:ParameterValuesCCZSTunableCoupler}}
\renewcommand{\arraystretch}{1.25}
\renewcommand{\tabcolsep}{0.15cm}
\begin{tabular}{| c | c c c c c c |}
\hline
$i$ & $\omega_i / 2\pi$ & $\alpha_i / 2\pi$ & $\omega_{c_j}^0 / 2\pi$ & $\alpha_{c_j} / 2\pi$ & $g_{i 1} / 2\pi$ & $g_{i 2} / 2\pi$ \\
\hline
0 & 4.8 & -0.17 & & & 0.07 & 0.07 \\
1 & 4.225 & -0.18 & 7.8 & -0.12 & 0.07 & \\
2 & 4.35 & - 0.18 & 8.0 & -0.12 & & 0.07 \\ 
\hline
\end{tabular}
\end{table}

We next consider the setup with tunable couplers as shown in \figref{fig:CCZSTunableCouplerSetup}. This setup is modelled with the Hamiltonian
\bea
H &=& \sum_{i = 0, 1, 2} \mleft[ \omega_i a_i^\dag a_i + \frac{\alpha_i}{2} a_i^\dag a_i \mleft( a_i^\dag a_i - 1 \mright) \mright] \nn \\
&&+ \sum_{j = 1, 2} \mleft[ \omega_{c_j} (t) b_j^\dag b_j + \frac{\alpha_{c_j}}{2} b_j^\dag b_j \mleft( b_j^\dag b_j - 1 \mright) \mright] \nn \\
&&+ \sum_{\substack{i = 0, 1, 2 \\ j = 1, 2}} g_{i j} \mleft( a_i^\dag + a_i \mright) \mleft( b_j^\dag + b_j \mright),
\label{eq:HTunableCoupler}
\eea
where $a_i$ and $a_i^\dag$ ($b_j$ and $b_j^\dag$) are the annihilation and creation operators, respectively, of qubit $i$ (coupler $j$), $\omega_i$ ($\omega_{c_j}$) is its transition frequency, $\alpha_i$ ($\alpha_{c_j}$) its anharmonicity, and $g_{ij}$ is the strength of the capacitive coupling between qubit $i$ and coupler $j$. We use parameter values similar to recent updates of the design in Ref.~\cite{Bengtsson2020}. These values, which are kept fixed throughout all simulations, are given in \tabref{tab:ParameterValuesCCZSTunableCoupler}. 

To activate the gate, the magnetic flux $\Phi_j (t)$ through the superconducting quantum interference device (SQUID) of coupler $j$ (see \figref{fig:CCZSTunableCouplerSetup}) is modulated as
\be
\Phi_j (t) = \Theta_j + \delta_j (t) \cos (\omega_{\Phi_j} t + \varphi_j),
\ee
where $\Theta_j$ is the DC bias, $\delta_j (t)$ is an envelope function with sinusoidal rise and fall of \unit[25]{ns} and a constant value $\delta_{0_j}$ for a time $t_p$ in-between such that $t_{\rm gate} = t_p + \unit[50]{ns}$, $\omega_{\Phi_j}$ is the modulation frequency [close to resonance with the transition frequency between the states that are coupled by the CZ gate (see \figref{fig:PrincipleCCZS})], and $\varphi_j$ is the initial phase of the drive, which is kept equal to zero until we need to calibrate different values of $\phi$ in the CCZS gate family. Modulating a symmetric SQUID like this results in a time-dependent coupler frequency~\cite{Koch2007}
\be
\omega_{c_j} (t) = \omega_{c_j}^0 \sqrt{\abs{\cos \mleft( \pi \frac{\Phi (t)}{\Phi_0} \mright)}},
\ee
where $\Phi_0$ is the flux quantum. Our control parameters for the CCZS gate are thus $t_p$, $\delta_{0_j}$, $\Theta_j$, and the detuning between $\omega_{\Phi_j}$ and the expected resonant frequency.


\paragraph{Calibration procedure and results}

\begin{figure*}[]
\includegraphics[width=0.9\linewidth]{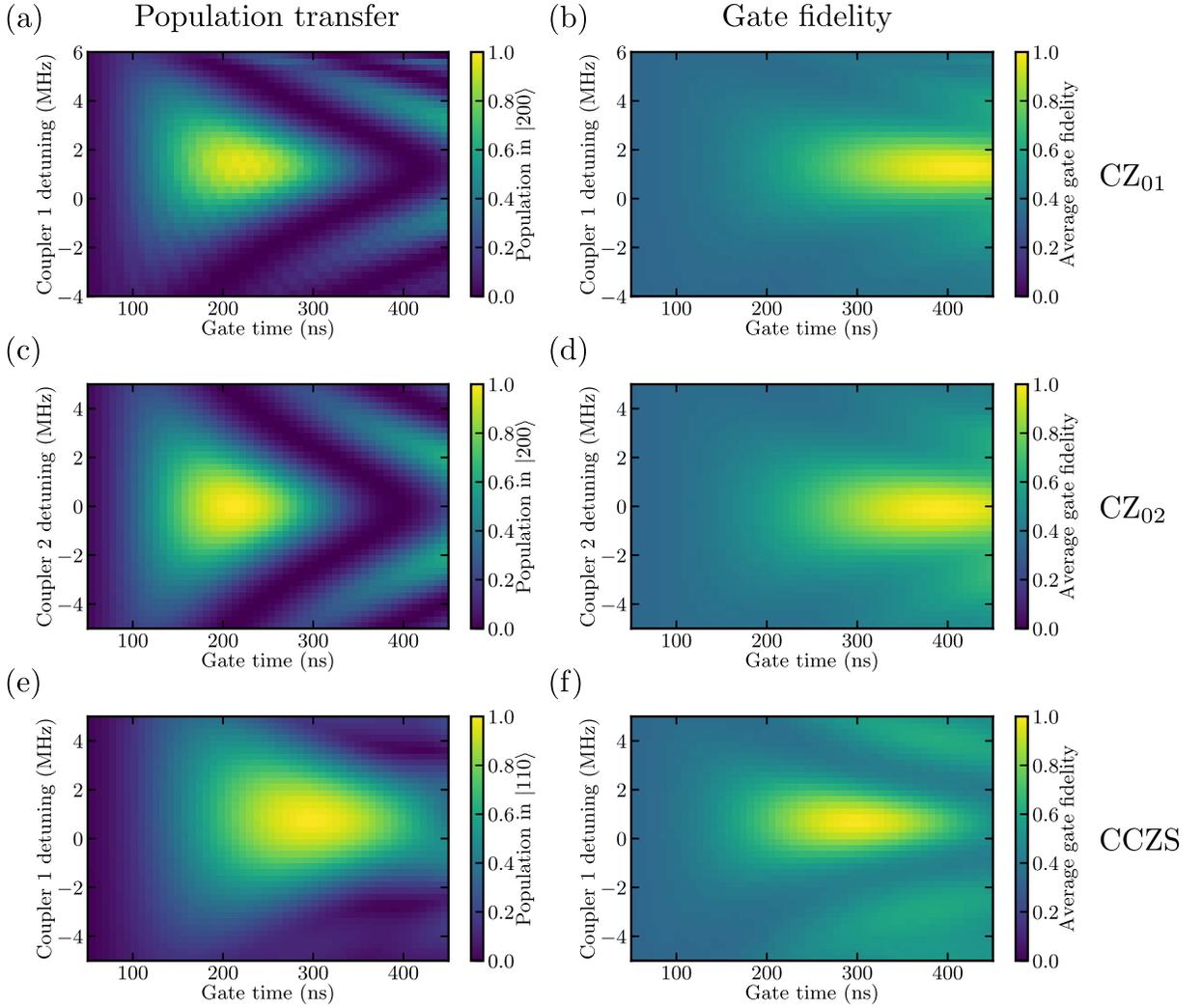}
\caption{Calibrating the CCZS gate with tunable couplers.
(a) Population in $\ket{200}$ as a function of gate time and the detuning between the modulation frequency $\omega_{\Phi_1}$ and the frequency of the transition $\ket{110} \leftrightarrow \ket{200}$ when calibrating the $\text{CZ}_{01}$ gate by initializing the system in $\ket{110}$.
(b) The gate fidelity for the $\text{CZ}_{01}$ gate for the parameters in (a).
(c) Population in $\ket{200}$ as a function of gate time and the detuning between the modulation frequency $\omega_{\Phi_2}$ and the frequency of the transition $\ket{101} \leftrightarrow \ket{200}$ when calibrating the $\text{CZ}_{02}$ gate by initializing the system in $\ket{101}$.
(d) The gate fidelity for the $\text{CZ}_{02}$ gate for the parameters in (c).
(e) Population in $\ket{110}$ as a function of gate time and the detuning between the modulation frequency $\omega_{\Phi_1}$ and the frequency of the transition $\ket{110} \leftrightarrow \ket{200}$ when calibrating the $\text{CCZS} (\pi / 2, \pi, 0)$ gate by initializing the system in $\ket{101}$. During the calibration, we also vary $\omega_{\Phi_2}$, but in this plot it is kept fixed.
(f) The gate fidelity for the $\text{CCZS} (\pi / 2, \pi, 0)$ gate for the parameters in (e).
\label{fig:CCZSTunableCouplerResults}}
\end{figure*}

We calibrate gates in the CCZS family by the following procedure:
\begin{enumerate}
\item We first tune up high-fidelity $\text{CZ}_{01}$ and $\text{CZ}_{02}$ gates with equal gate times. Both CZ gates must be implemented such that the second excited state used is that of qubit 0, as shown in \figpanel{fig:PrincipleCCZS}{b}. We begin by exploring the parameter space spanned by $\Theta_j$ and $\delta_{0_j}$ to find values that yield high population transfers from $\ket{1_0 1_j}$ to $\ket{2_0 0_j}$. We then plot the population in $\ket{2_0 0_j}$ as a function of $t_{\rm gate}$ and $\omega_{\Phi_j}$ as in \figpanel{fig:CCZSTunableCouplerResults}{a} and \figpanelNoPrefix{fig:CCZSTunableCouplerResults}{c}  and go along the value of $\omega_{\Phi_j}$ that corresponds to the tip of the resulting chevron pattern to the first value of $t_{\rm gate}$ that returns all population to $\ket{1_0 1_j}$. Finally, we confirm that the CZ gate fidelity around this point in parameter space is close to \unit[100]{\%} and pick the parameter values in this area that give the highest gate fidelity.
\item Next, we apply pulses to both couplers with the same DC biases and amplitudes as for the good CZ gates found in the previous step, but sweep the modulation frequency of both pulses around the values for the CZ gates. The smoking gun for the CCZS gates is a maximal population transfer between $\ket{101}$ and $\ket{110}$, which corresponds to $\theta = \pi / 2$. We expect such a point in parameter space to show up at gate times around $\sqrt{2}$ shorter than those found for the CZ gates. Having found such a point, we check that the gate fidelity around that point is close to \unit[100]{\%} for the $\text{CCZS} (\pi / 2, \phi, 0)$ gate. We then pick the parameter values in this area that give the highest gate fidelity for the desired value of $\phi$.
\item Other elements of the CCZS gate family can also be found, but none of them have such a clear signature as the $\ket{101} \leftrightarrow \ket{110}$ population transfer. In particular, other values for the phase $\phi$ are found by changing the relative initial phase between the two pulses, $(\varphi_1 - \varphi_2)$. Other values of $\theta$ can, in principle, be found by combining the controls of individual CZ gates with different gate strengths and adjusting the gate time accordingly [see the discussion below \eqref{eq:UCZS_gamma0}].
\end{enumerate}

We now show how such a calibration procedure can look like in practice. We first plot the population in $\ket{2_0 0_j}$ as a function of $t_{\rm gate}$ and $\omega_{\Phi_j}$ for the two individual CZ gates in \figpanel{fig:CCZSTunableCouplerResults}{a} and \figpanelNoPrefix{fig:CCZSTunableCouplerResults}{c}. We find high-fidelity ($> \unit[99.7]{\%}$ and $> \unit[99.9]{\%}$) CZ gates with similar gate times, around 405 and \unit[396]{ns}, by choosing a DC bias $\Theta_j = 0.275 \Phi_0$ for both couplers and an amplitude $\delta_{0_j} \approx 0.08 \Phi_0$. In \figpanel{fig:CCZSTunableCouplerResults}{b} and \figpanelNoPrefix{fig:CCZSTunableCouplerResults}{d}, we show the corresponding maps of gate fidelity as a function of gate time and $\omega_{\Phi_j}$.

We then try applying the same pulses simultaneously. The parameters we vary are now the two modulation frequencies (and the gate time), so we need to look at different projections of the resulting 2-dimensional parameter space. In \figpanel{fig:CCZSTunableCouplerResults}{e}, we show one such projection, fixing $\omega_{\Phi_2}$ and plotting the population in $\ket{101}$ as a function of $\omega_{\Phi_1}$ and $t_{\rm gate}$. The corresponding gate fidelity for $\text{CCZS} (\pi / 2, \pi, 0)$ as a function of the same parameters is plotted in \figpanel{fig:CCZSTunableCouplerResults}{f}. Selecting the parameters that yield the highest population transfer, and optimizing for fidelity around those values, we find a $\text{CCZS} (\pi / 2, \pi, 0)$ gate with $t_{\rm gate} = \unit[295]{ns}$, which is a factor $\sim \sqrt{2}$ shorter plateau time $t_p$ than the individual CZ gates. The gate fidelity is $>\unit[99.3]{\%}$.

Note that, just as for the simulations with tunable qubits above, we have not included any effects of decoherence in these simulations. The impact of decoherence will be less the faster the gates are. It is possible to calibrate faster CZ and CCZS gates than the examples shown here in \figref{fig:CCZSTunableCouplerResults} by choosing other values of $\Theta_j$ and $\delta_{0_j}$, but we have chosen to show these examples since they illustrate the calibration and workings of the gates more clearly than some of the faster gates.


\paragraph{Tuning the gate parameter $\phi$}

\begin{figure}[]
\includegraphics[width=\linewidth]{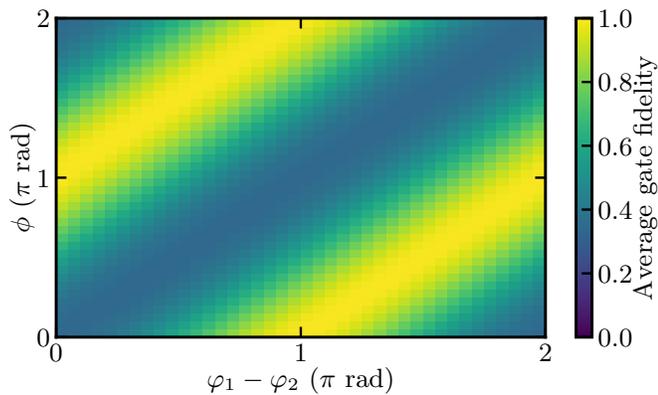}
\caption{Calibrating the $\text{CCZS} (\pi / 2, \phi, 0)$ gate by tuning the phase difference $\varphi_1 - \varphi_2$ between the signals modulating the two tunable couplers. The plots shows the gate fidelity for the $\text{CCZS} (\pi / 2, \phi, 0)$ gate as a function of $\phi$ and $\varphi_1 - \varphi_2$. All other parameters are the same as those that gave the highest gate fidelity for the $\text{CCZS} (\pi / 2, \pi, 0)$ in \figref{fig:CCZSTunableCouplerResults}. For each value of $\phi$ shown in the plot, the highest gate fidelity exceeds \unit[99]{\%}.
\label{fig:TuningPhi}}
\end{figure}

To further demonstrate the extensive control of gate parameters afforded by the setup with tunable couplers, we calibrate the $\text{CCZS} (\pi / 2, \phi, 0)$ gate for many values of $\phi$ in the range $[0 , 2 \pi]$. We do this by starting from the optimized parameters for $\text{CCZS} (\pi / 2, \pi, 0)$ found above and then tuning the phase difference $\varphi_1 - \varphi_2$ between the signals modulating the two tunable couplers. The resulting gate fidelities for $\text{CCZS} (\pi / 2, \phi, 0)$ are shown in \figref{fig:TuningPhi}. For all values of $\phi$ we try, we find gate fidelities above \unit[99]{\%}. These high gate fidelities are achieved along the line $\phi = \pi + \varphi_1 - \varphi_2 \mod{2 \pi}$, as expected from \eqref{eq:ThetaPhiRelation}.


\paragraph{Error sources}

Just as for the setup with tunable qubits in \secref{sec:ExpCCZSTunableCouplers}, we do not reach perfect \unit[100]{\%} gate fidelity in our simulations of the setup with tunable couplers, despite neglecting decoherence effects. The remaining error has multiple contributions. Firstly, the pulse shape $\delta_j (t)$ is chosen to be very simple; no optimal control is applied to improve it. Secondly, higher-order interactions between the qubits mediated by the off-resonant couplers result in ZZ interactions that disturb the three-qubit gate. We observe higher gate fidelities if we allow ourselves to correct phases like those produced by such interactions. This suggests that schemes for reducing unwanted ZZ interactions in two-qubit gates (see, e.g., Refs.~\cite{Foxen2020, Sung2021}) could be helpful also for the three-qubit gate considered here.

Thirdly, we note that we restricted ourselves to calibrating gates on the form $\text{CCZS} (\pi / 2, \phi, 0)$. It is possible that some of the gates we produced had higher gate fidelities with $\text{CCZS} (\theta, \phi, \gamma)$ for other values of $\theta$ and $\gamma$, but we preferred tuning up and showing fidelities for a gate with clearer functionality rather than searching the space of parameters $\theta$ and $\gamma$ to find the highest possible fidelity. Finally, the simulations with five three-level transmon qubits are quite computationally heavy and we needed to search a 10-dimensional parameter space (plateau times $t_p$, modulation frequencies $\omega_{\Phi_j}$, modulation phases $\phi_j$, modulation amplitudes $\delta_{0_j}$, and DC biases $\Theta_j$). There is thus certainly room for improvement in exploration of this parameter space.


\section{Simultaneous iSWAP gates}
\label{sec:SimultaneousiSWAP}

In this section, we show that the idea of applying simultaneous two-qubit gates to create multi-qubit gates is not limited to the CZ gates studied in \secref{sec:SimultaneousCZ}. Here, we investigate what happens when the simultaneous gates are iSWAP gates instead. The treatment in this section will be more condensed than in the previous one, since some parts turn out to be quite similar. We note that several other combinations of two-qubit gates are possible, but the detailed study of such possibilities is left for future work.


\subsection{Setup and gate operation}

We consider simultaneous application of iSWAP gates that are based on activating a coupling between the states $\ket{01}$ and $\ket{10}$. In such gates, the states $\ket{00}$ and $\ket{11}$ do not couple to other states and remain unchanged while the states $\ket{01}$ and $\ket{10}$ are swapped and acquire a phase factor $-i$. In superconducting circuits, just as for the CZ gate, the required coupling can be achieved either by tuning the frequencies of the two qubits into resonance~\cite{Bialczak2010, Barends2019}, possibly in conjunction with tuning a coupler~\cite{Foxen2020, Sung2021}, by parametrically modulating a tunable coupler between two fixed-frequency qubits~\cite{McKay2016, Ganzhorn2019, Ganzhorn2020}, or by parametrically modulating one of the qubits~\cite{Caldwell2018, Abrams2020}. In Ref.~\cite{Sung2021}, a gate fidelity of \unit[99.86]{\%} was reported for a gate time of $\unit[30]{ns}$.


\subsubsection{Hamiltonians and time evolution}

\begin{figure}[]
\includegraphics[width=\linewidth]{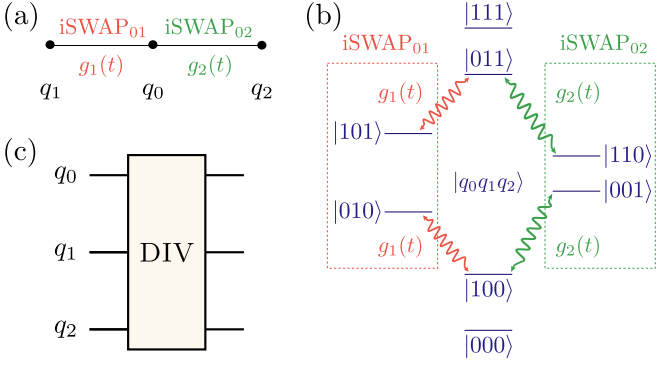}
\caption{Setup and operation for the three-qubit gate realized through simultaneous application of two iSWAP gates.
(a) The setup considered is a linear chain of three qubits with nearest-neighbour coupling. The iSWAP gates $\mathrm{iSWAP}_{0j}$ between qubits $0$ and $j = \{ 1, 2 \}$ are applied simultaneously by activating couplings between the $\ket{1_0 0_j}$ and $\ket{0_0 1_j}$ states with the coupling strength $g_j$.
(b) The transitions in the three-qubit system activated by the application of the iSWAP gates. The states $\ket{000}$ and $\ket{111}$ are not affected by the gates.
(c) We denote the three-qubit operation resulting from the simultaneous application of the two iSWAP gates by DIV, since it is a ``divider'' gate that can distribute one or two excitations among all three qubits.
\label{fig:PrincipleSimultaneousiSWAP}}
\end{figure}

We consider the same linear chain of three qubits as for the CCZS gate in \figref{fig:PrincipleCCZS}, but now coupling the transitions between the states $\ket{1_0 0_j}$ and $\ket{0_0 1_j}$ with strengths $g_j$, as illustrated in  \figpanel{fig:PrincipleSimultaneousiSWAP}{a}. A simplification compared to the simultaneous CZ gates is that no second excited state of any qubit becomes part of the dynamics. For the simultaneous iSWAP gates, only the transitions shown in \figpanel{fig:PrincipleSimultaneousiSWAP}{b} are activated.

Assuming for simplicity that the transitions shown in \figpanel{fig:PrincipleSimultaneousiSWAP}{b} are resonant ($\delta = 0$), and that all other transitions are far off resonance, the Hamiltonian for the system can in the interaction picture be written as
\be
H = \sigma_0^{10} \mleft( g_1 \sigma_1^{01} + g_2 \sigma_2^{01} \mright) + \text{H.c.},
\label{eq:HamiltonianSimultaneousiSWAP}
\ee
with operators defined as in \secref{sec:DickeHamiltonianAndDynamics}. This Hamiltonian conserves the number of excitations in the system and we also see that the transitions in \figpanel{fig:PrincipleSimultaneousiSWAP}{b} occur within the subspaces determined by the excitation number.

In the subspaces with one and two excitations, $\Lambda$ or $V$ systems are formed. Just as in \secref{sec:SimultaneousCZHamiltonians}, it is thus convenient to introduce new basis states that include dark and bright states. Defining
\bea
\ket{B_1} &=& \frac{g_1 \ket{10} + g_2 \ket{01}}{\Omega}, \\
\ket{B_2} &=& \frac{g_2 \ket{10} + g_1 \ket{01}}{\Omega}, \\
\ket{D_1} &=& \frac{g_2 \ket{10} - g_1 \ket{01}}{\Omega}, \\
\ket{D_2} &=& \frac{g_1 \ket{10} - g_2 \ket{01}}{\Omega},
\eea
where
\be
\Omega = \sqrt{g_1^2 + g_2^2},
\ee
the unitary dynamics generated by the system Hamiltonian in \eqref{eq:HamiltonianSimultaneousiSWAP} can be expressed as
\bea
U &=& \ketbra{000}{000} + \ketbra{111}{111} \nn \\
&&+ \ketbra{0 D_2}{0 D_2} + \ketbra{1 D_1}{1 D_1} \nn \\
&&+ \cos \mleft( \Omega t \mright) I' - i \sin \mleft( \Omega t \mright) \sigma_x' \nn \\
&&+ \cos \mleft( \Omega t \mright) I'' - i \sin \mleft( \Omega t \mright) \sigma_x'',
\label{eq:USimiSWAP}
\eea
where $I'$ and $\sigma_x'$ are defined in the subspace spanned by $\ket{100}$ and $\ket{0 B_1}$, and $I''$ and $\sigma_x''$ are defined in the subspace spanned by $\ket{011}$ and $\ket{1 B_2}$.


\subsubsection{The family of three-qubit gates}

The results above show that the states in the computational subspace of the three qubits are affected as follows: $\ket{000}$ and $\ket{111}$ are unchanged, while states are swapped around in the single- and double-excitation subspaces. By introducing the notation $\tan \theta =  g_2 / g_1$ (we assume for simplicity that $g_1$ and $g_2$ are in phase) and $\varphi = \Omega t$, we can write the three-qubit gate, which we denote DIV [see \figpanel{fig:PrincipleSimultaneousiSWAP}{c}], as
\be
U_{\rm DIV} (\theta, \varphi) = U_0 \oplus U_1 (\theta, \varphi) \oplus U_2 (\theta, \varphi) \oplus U_3,
\ee
where $U_j$ acts on the $j$-excitation subspace. Here, $U_0 = U_3 = 1$, while $U_1$ and $U_2$ are found by transforming from the basis with bright and dark states used in \eqref{eq:USimiSWAP} (see \appref{app:CZ}) to the computational basis; in the single-excitation subspace spanned by $\ket{010}$, $\ket{100}$, and $\ket{001}$, we obtain
\bea
&& U_1 (\theta, \varphi) = \\
&&\begin{bmatrix}
\sin^2 \theta + \cos^2 \theta \cos \varphi & - i \cos \theta \sin \varphi & \frac{1}{2} \sin 2 \theta \mleft( \cos \varphi - 1 \mright) \\
- i \cos \theta \sin \varphi & \cos \varphi & - i \sin \theta \sin \varphi \\
\frac{1}{2} \sin 2 \theta \mleft( \cos \varphi - 1 \mright) & - i \sin \theta \sin \varphi & \cos^2 \theta + \sin^2 \theta \cos \varphi
\end{bmatrix} \nn
\label{eq:U1}
\eea
and $U_2 (\theta, \varphi)$ for the double-excitation subspace spanned by $\ket{101}$, $\ket{011}$, and $\ket{110}$ has exactly the same form.

An important difference between the DIV gate from simultaneous iSWAP gates and the CCZS gate from simultaneous CZ gates in the previous section is that the operation of the DIV gate never makes any population leave the computational subspace. This is why we can vary the parameter $\varphi$ freely by choosing the evolution time $t$. In the CCZS gate, the evolution time is heavily constrained by the need to ensure that the temporary population in the middle qubit's second excited state returns to the computational subspace at the end of the gate.


\subsubsection{Examples of three-qubit gates}
\label{sec:ExamplesDIV}

We now study some simple parameter choices for the DIV gate. If we set $g_1 = g$, $g_2 = 0$, and $\varphi = \pi / 2$, i.e., $t_{\rm gate} = \pi / 2 g$, we recover the two-qubit iSWAP gate acting on qubits 0 and 1. In the same way, if $g_1 = 0$, $g_2 = g$, and $\varphi = \pi / 2$, we have the two-qubit iSWAP gate acting on qubits 0 and 2.

If we activate both these iSWAP interactions simultaneously, i.e., $g_1 = g_2 = g$ such that $\theta = \pi / 4$, and choose the gate time $t_{\rm gate} = \pi / 2 \sqrt{2} g$ such that $\varphi = \pi / 2$, we obtain $\text{DIV} (\pi / 4, \pi / 2)$, for which
\be
U_{1, 2} (\pi / 4, \pi / 2) =
\begin{bmatrix}
\frac{1}{2} & - \frac{i}{\sqrt{2}} & - \frac{1}{2} \\
- \frac{i}{\sqrt{2}} & 0 & - \frac{i}{\sqrt{2}} \\
- \frac{1}{2} & - \frac{i}{\sqrt{2}} & \frac{1}{2}
\end{bmatrix}.
\label{eq:U1Simplest}
\ee
This gate, which is a factor $\sqrt{2}$ faster than the individual two-qubit iSWAP gates, thus takes a single excitation in the middle qubit and divides it evenly between the two outer qubits. A single excitation in one of the outer qubits ends up divided across all three qubits: half in the middle qubit and a quarter in each of the outer qubits.

If we keep the two coupling strengths the same ($g_1 = g_2 = g$ such that $\theta = \pi / 4$), but vary the parameter $\varphi$ by varying the gate time $t = \varphi / \sqrt{2} g$, the gate becomes
\be
U_{1, 2} (\pi / 4, \varphi) = 
\begin{bmatrix}
\frac{1}{2} \mleft( 1 + \cos \varphi \mright) & - \frac{i}{\sqrt{2}} \sin \varphi & \frac{1}{2} \mleft( \cos \varphi - 1 \mright) \\
- \frac{i}{\sqrt{2}} \sin \varphi & \cos \varphi & - \frac{i}{\sqrt{2}} \sin \varphi \\
\frac{1}{2} \mleft( \cos \varphi - 1 \mright) & - \frac{i}{\sqrt{2}} \sin \varphi & \frac{1}{2} \mleft( 1 + \cos \varphi \mright)
\end{bmatrix}.
\label{eq:U1ThetaPi4}
\ee
If we instead fix $\varphi = \pi / 2$, but vary $\theta$ by varying the ratio of the coupling strengths $g_1$ and $g_2$, the resulting gate is given by
\be
U_{1, 2} (\theta, \pi / 2) =
\begin{bmatrix}
\sin^2 \theta & - i \cos \theta & - \frac{1}{2} \sin 2 \theta \\
- i \cos \theta & 0 & - i \sin \theta \\
- \frac{1}{2} \sin 2 \theta & - i \sin \theta & \cos^2 \theta
\end{bmatrix}.
\label{eq:U1VarphiPi2}
\ee
%


\subsection{Decomposition into two-qubit gates}

We note that all the three-qubit gates in \secref{sec:ExamplesDIV} entangle all three qubits. Although finding a decomposition of the DIV gate into single- and two-qubit gates is less straightforward than for the CCZS gate in \secref{sec:DecompositionCCZS}, we can from this entanglement conclude that at the very least two sequential two-qubit gates are necessary for such a decomposition. Since the three-qubit gate already is faster than a single two-qubit gate, this guarantees a significant speed-up.


\subsection{Constructing other three-qubit gates}

Unlike the CCZS gate, the DIV gate cannot be interpreted as one qubit controlling a two-qubit operation on the other two qubits. The most well-known three-qubit gates, the Fredkin, iFredkin, and Toffoli gates considered in \secref{sec:CCZSOther3qGates}, are all such gates. It is thus clear that the DIV gate cannot be equivalent to any of these three-qubit gates for any choice of parameters $\theta$ and $\varphi$. Furthermore, it does not appear possible to change any DIV gate into such a form by adding a single two-qubit gate before or after the DIV gate.


\subsection{Creating large entangled states}
\label{sec:DIVEntanglement}

\begin{figure}[]
\includegraphics[width=0.9\linewidth]{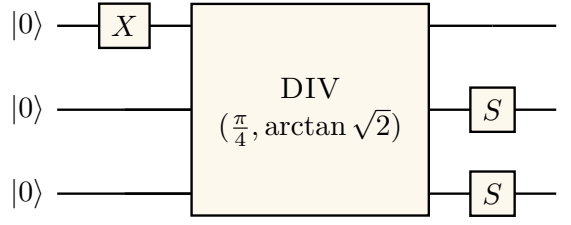}
\caption{Quantum circuit for generating the three-qubit W state $\mleft( \ket{100} + \ket{010} + \ket{001} \mright) / \sqrt{3}$ using the DIV gate. The phase gate S is $\sqrt{\text{Z}}$.
\label{fig:WDIV}}
\end{figure}

We now turn to how the DIV gate and its generalizations to more qubits can be used to rapidly create large entangled states, similar to what we showed for the CCZS gate in \secref{sec:CCZSEntanglement}. We first note that arbitrary superpositions of all permutations of three-qubit states with one excitation can easily be created by starting from $\ket{000}$, exciting qubit 0, and then applying the DIV gate for suitable values of $\theta$ and $\varphi$, yielding
\bea
&&U_1(\theta, \varphi) \ket{100} = \nn \\
&&\cos \varphi \ket{100} - i \sin \varphi \mleft( \cos \theta \ket{010} + \sin \theta \ket{001} \mright).
\label{eq:ArbitraryWDIV} 
\eea
In particular, the three-qubit W state
\be
\ket{D_3^1} = \frac{1}{\sqrt{3}} \mleft( \ket{100} + \ket{010} + \ket{001} \mright)
\ee
can be constructed by choosing $\theta = \pi / 4$ (i.e., $g_1 = g_2$) and $\varphi = \arctan \sqrt{2}$ in \eqref{eq:ArbitraryWDIV}, and following that by applying single-qubit gates to qubits 1 and 2, as shown in \figref{fig:WDIV}.

We note that the experiment in Ref.~\cite{Neeley2010} showed how to construct a W state by single-qubit rotations and a single application of simultaneous iSWAP interactions between all three qubits in a triangular setup instead of the linear chain considered here. A generalization of this protocol to more qubits is given in Ref.~\cite{Neeley2010a}. Furthermore, a protocol to construct the three-qubit GHZ state using single-qubit rotations and a single application of simultaneous iSWAP interactions in a linear chain like we consider here was presented in Ref.~\cite{Galiautdinov2008}.

The multi-qubit version of the simultaneous iSWAP interaction, where the dynamics of a central qubit 0 and its $N$ nearest neghbours $\{ j \}$ are governed by the interaction-picture Hamiltonian
\be
H = \sigma_0^{10} \sum_{j = 1}^N g_j \sigma_j^{01} + \text{H.c.},
\label{eq:HNiSWAP}
\ee
is the same as the simultaneous CZ interaction given in \eqref{eq:HNCZ} except that it is the $\ket{0} \leftrightarrow \ket{1}$ transition that couples to the surrounding qubits instead of the $\ket{1} \leftrightarrow \ket{2}$ transition. We can thus reuse much of what we derived in \secref{sec:DickeCCZS} about how to rapidly create large Dicke states. For example, the procedure for creating W states with many qubits described in \figref{fig:WStatesCZScalingUp} can also be implemented with simultaneous iSWAP gates. It is actually even easier, since the initial qubit only needs to be prepared in state $\ket{1}$ instead of $\ket{2}$ and the later step of flipping $\ket{1}$ to $\ket{2}$ for other qubits described there can be omitted.

To create superpositions of Dicke states like in \secref{sec:5QubitDicke}, only a minor modification of the protocol presented there is needed to adapt it to simultaneous iSWAP gates. We simply change the states $\ket{0_0}$, $\ket{1_0}$, and $\ket{2_0}$ to $\ket{2_0}$, $\ket{0_0}$, and $\ket{1_0}$, respectively, during the execution of the protocol. At the end, we change them back to obtain the state in \eqref{eq:D53}.


\subsection{Experimental feasibility}
\label{sec:ExpiSWAP}

In the same way as for the CCZS gate (see \secref{sec:ExpCCZS}), there are several experimental setups with superconducting qubits that could be used to implement the simultaneous iSWAP gates that make up the DIV gate. This includes setups with tunable qubits, where the states $\ket{01}$ and $\ket{10}$ are tuned into resonance~\cite{Bialczak2010, Barends2019}. This activation of the iSWAP gate can be further enhanced with a tunable coupler~\cite{Foxen2020, Sung2021}. The other type of setup uses parametric modulation of either a tunable coupler~\cite{McKay2016, Ganzhorn2019, Ganzhorn2020} or one of the qubits~\cite{Caldwell2018, Abrams2020}.

For brevity and simplicity, we here limit our simulations to the implementation with tunable qubits. In this implementation, the parameter $\theta$ is fixed by the coupling strengths set in the hardware and cannot be changed in experiment. An implementation with parametric modulation of tunable couplers instead would enable controlling $\theta$ in situ. Such an implementation can be calibrated in similar fashion as the CCZS gate with tunable couplers in \secref{sec:ExpCCZSTunableCouplers}.

\begin{figure}[]
\includegraphics[width=\linewidth]{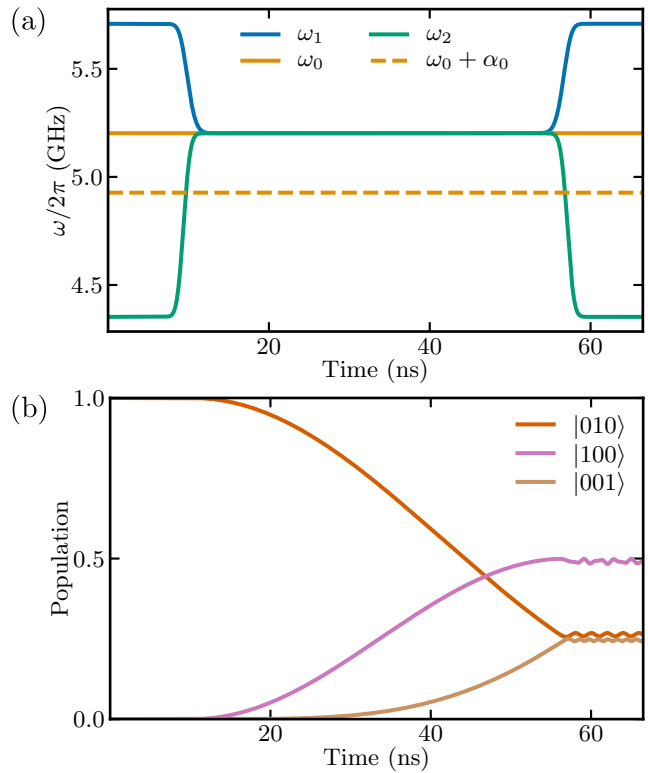}
\caption{Implementing the $\mathrm{DIV} (\pi / 4, \pi / 2)$ gate with tunable qubits. 
(a) The tuning of the qubit energies used to realize the gate.
(b) Population of the states $\ket{100}$, $\ket{010}$, and $\ket{001}$ (the single-excitation subspace) as a function of time during the gate when the initial state is $\ket{010}$. As \eqref{eq:U1Simplest} shows, the effect of the gate is to divide an excitation in one of the outer qubits such that half ends up in the middle qubit and a quarter each in the outer qubits.
\label{fig:DIVTunableQubits}}
\end{figure}

For the implementation with tunable qubits, we consider the same setup and parameters as in \secref{sec:ExpCCZSTunableQubits} except that we increase the maximum energy of qubit 2 to $\omega_0$; see \figpanel{fig:TunableQubitsCCZS}{a} and Eqs.~(\ref{eq:OmegaValuesCCZSTunableQubits})--(\ref{eq:LambdaValuesCCZSTunableQubits}) with $g_j = \lambda_j / 2$. We further use the same pulse shapes as in \figpanel{fig:TunableQubitsCCZS}{b}, but now tuning $\omega_1$ and $\omega_2$ into resonance with $\omega_0$ instead of $\omega_0 + \alpha_0$ and adapting the gate times to yield the iSWAP and DIV gates, resulting in the tuning shown in \figpanel{fig:DIVTunableQubits}{a}.

We first tune up the individual $\text{iSWAP}_{01}$ and $\text{iSWAP}_{02}$ gates by tuning just one of the outer qubits into resonance with the middle qubit. For the $\text{iSWAP}_{01}$ gate, we find a gate fidelity of \unit[99.8]{\%} with a gate time $t_{\rm gate, iSWAP_{01}} = \unit[66.8]{ns}$, and for the $\text{iSWAP}_{02}$ gate, we find a gate fidelity of \unit[99.6]{\%} using the same gate time.

We then tune up the $\text{DIV} ( \pi / 4, \pi / 2)$ gate [see \eqref{eq:U1Simplest}] by tuning both outer qubits into resonance with the middle qubit in a synchronized fashion, reducing the gate time by around a factor $\sqrt{2}$. Note that other values of $\varphi$ are easily implemented by decreasing or increasing the gate time, while $\theta = \pi / 4$ is fixed by the choice of $g_1 = g_2$ used here. For the $\text{DIV} ( \pi / 4, \pi / 2)$ gate, we find a gate fidelity of \unit[99.1]{\%} for the gate time $t_{\rm gate, DIV} = \unit[47.5]{ns} \approx t_{\rm gate, iSWAP} / \sqrt{2}$. Just as in \secref{sec:ExpCCZS}, the gate fidelity is calculated without including any effects of decoherence in the simulation. To illustrate the calibrated gate, we plot in \figpanel{fig:DIVTunableQubits}{b} the population transfers in the single-excitation subspace when the system is initialized in the state $\ket{010}$.

The deviation from perfect gate fidelity can be attributed to several factors, similar to the case of the CCZS gate with tunable qubits in \secref{sec:ExpCCZSTunableQubits}. As we tune qubit 2 into resonance with qubit 0, we pass the point where $\omega_2 = \omega_0 + \alpha_0$, i.e., where the $\ket{1_0 1_2} \leftrightarrow \ket{2_0 0_2}$ transition is resonant. This can cause leakage from the computational subspace. We note that qubit 1 does not have the same issue, which likely helps explain why we find a higher gate fidelity for the $\text{iSWAP}_{01}$ gate than the $\text{iSWAP}_{02}$. Just like for the CCZS gate, it would likely be beneficial to apply more advanced methods for optimizing the tuning of the qubits in and out of resonance. 


\section{Conclusion}
\label{sec:Conclusion}

We have shown how multi-qubit gates can be constructed and implemented by simultaneously applying two-qubit gates to a group of qubits such that at least one qubit is affected by the operation of two or more of these two-qubit gates. The resulting multi-qubit gates are as fast as, and in many cases clearly faster than, the individual two-qubit gates on their own. Furthermore, the multi-qubit gates can have larger entangling power than the sequential application of the constituent two-qubit gates, in addition to being much faster than such a sequential application.

Since our scheme for multi-qubit gates only relies on control operations corresponding to two-qubit gates, our ideas are ready to be implemented in existing quantum hardware without the need for any additional components, complicated pulse shapes, hardware re-design, or other changes beyond some recalibration of the lengths (and in some cases, the phases) of the control pulses already optimized for two-qubit gates. This means that the multi-qubit gates presented in this article, and other multi-qubit gates using the same principles, could become useful immediately across mature quantum-computing platforms like superconducting circuits, trapped ions, and others.

We illustrated our ideas for multi-qubit gates with two specific examples: simultaneously applied interactions for CZ gates and simultaneously applied interactions for iSWAP gates. For the simultaneous CZ gates, based on activating the $\ket{11} \leftrightarrow \ket{02}$ transition, we showed that applying them to the nearest neighbours in a linear chain of three qubits, with the middle qubit being the one excited to $\ket{2}$, resulted in a three-qubit gate that we denoted CCZS. This CCZS gate applies a combination of the CZ and SWAP gates to the outer qubits in the chain conditioned on the middle qubit being in its excited state. By controlling the ratio of amplitudes and the relative phases of the pulses for the constituent CZ gates, and also the detuning from resonance of the $\ket{11} \leftrightarrow \ket{02}$ transition, we gain access to a whole family of CCZS gates. For the case where the CZ control pulses are in phase, on resonance, and of equal amplitude, the gate time for the CCZS gate is a factor $\sqrt{2}$ shorter than for a single CZ gate.

Exploring the entangling power of the CCZS gates, we showed that a decomposition of a gate in the CCZS family in general requires three sequential two-qubit gates. We also demonstrated that gates from the CCZS family can be used to construct other three-qubit gates: the iFredkin and Fredkin gates are equivalent to a CCZS gate followed by a two-qubit CZ gate or a three-qubit CCZ gate, respectively. For the iFredkin gate, this suggests that we can implement it twice as fast using a construction with a CCZS gate than a standard decomposition into two-qubit gates. Furthermore, we showed that a single CCZS gate combined with a few single-qubit gates can be used to construct an entangled three-qubit GHZ state. Finally, we generalized the CCZS gate operation to more qubits and showed that, in combination with a few single-qubit gates, it can create large entangled Dicke states in very few steps.

For the simultaneous iSWAP gates, based on activating the $\ket{01} \leftrightarrow \ket{10}$ transition, we showed that when they are applied in a linear chain of three qubits, a three-qubit gate which we denoted DIV is created. The DIV gate distributes excitations among the three qubits within the one- and two-excitation subspaces while leaving the states $\ket{000}$ and $\ket{111}$ unchanged. We showed that we can create a large family of DIV gates by controlling the gate time and the relative strength of the two constituent iSWAP gates. Furthermore, similar to the CCZS gates, we showed that the DIV gates are in general faster than single iSWAP gates and can be used to rapidly construct large entangled states like Dicke and GHZ states.

For both the CCZS gate and the DIV gate, we performed numerical simulations using parameters from existing state-of-the-art quantum hardware with superconducting qubits to demonstrate that these three-qubit gates are ready to be implemented with high fidelity in experiments. For the CCZS gate, we showed that it can be implemented with both tunable qubits and with tunable couplers, where the latter gives some more freedom to control parameters and realize the whole family of gates. We found that both setups enable gate fidelities exceeding $\unit[99.3]{\%}$, with the tunable qubits reaching a gate fidelity above $\unit[99.4]{\%}$. For the DIV gate, we limited the simulations to the setup with tunable qubits and demonstrated a gate fidelity exceeding $\unit[99.1]{\%}$. We emphasize that all these simulations used quite simple and straightforward methods for optimization and calibration, indicating that these high gate fidelities should be within reach in experiment. Furthermore, we identified factors contributing to the deviations from perfect gate fidelity, e.g., lack of optimal control applied to gate parameters varying in time and the presence of unwanted ZZ coupling. This allowed us to suggest several improvements to the operation of the three-qubit gates, which should enable even higher gate fidelities than demonstrated here.

In conclusion, we have introduced a general method for creating multi-qubit gates using two-qubit gates already in use in current quantum hardware. We have shown that these multi-qubit gates are fast, powerful, and ready to be implemented in existing experimental setups without any significant modifications needed. This opens up a wealth of possible applications by making quantum circuits more compact and faster to run, which is crucial for unleashing the potential of NISQ devices that are limited by coherence times.


\section{Outlook}
\label{sec:Outlook}

We see at least five directions for further research building on the results presented in this article. The first is to test the ideas detailed here in actual experiments. Such experiments could be performed using various setups with superconducting qubits, as we have analyzed in Sections~\ref{sec:ExpCCZS} and \ref{sec:ExpiSWAP}, but also on other quantum hardware platforms. We note that experimental implementations would benefit from further developing calibrations methods compared to what we showed in Sections~\ref{sec:ExpCCZS} and \ref{sec:ExpiSWAP}. In the numerical simulations there, we had access to the full propagator associated with the gate. This allowed us to simplify the calibration process substantially, since we could easily check the average gate fidelity for those points in parameter space that showed population transfers between the computational states corresponding to the gate we sought.

The question of experiments ties into the second research direction, which is to analyze how well the schemes from this article will perform on other platforms than superconducting qubits. Furthermore, it should be investigated whether there are two-qubit gate implementations native to these other platforms that can be run simultaneously to create new multi-qubit gates.

This last part can also be viewed as part of the third research direction, which is to find more multi-qubit gates realized through simultaneous application of other two-qubit gates than CZ and iSWAP, which were used as examples in this article. Candidates for such two-qubit gates include Controlled-NOT (CNOT) gates implemented through cross-resonance driving~\cite{Rigetti2010, Chow2011}. It may also be possible to simultaneously apply different two-qubit gates to different pairs of qubits to create yet other multi-qubit gates.

The fourth research direction we envision is to compile or transpile various quantum algorithms anew with the novel multi-qubit gates included in the native gate set of the device that the algorithm is to be executed on. We expect this to lead to a significantly reduced circuit depth and run time for some algorithms. As mentioned in the introduction, the CCZS gate seems particularly suited to improve phase estimation and spectrum qubitization, but there are likely many more algorithms that would benefit from its inclusion. For example, one could investigate the use of our multi-qubit gates in the entangling layers of variational quantum algorithms~\cite{Cerezo2020} like the quantum approximate optimization algorithm~\cite{Farhi2014}, the quantum alternating operator ansatz~\cite{Hadfield2019}, or the variational quantum eigensolver~\cite{Peruzzo2014}.

Finally, we also believe that tools from optimal control and insights from other works on optimizing pulse shaping for gates should be applied to the multi-qubit gates developed here. This could help achieve even higher gate fidelities and shorter circuit run times by reducing leakage to states outside the computational subspace and further decreasing the gate time.


\begin{acknowledgments}

We acknowledge support from the Knut and Alice Wallenberg Foundation through the Wallenberg Centre for Quantum Technology (WACQT) and from the EU Flagship on Quantum Technology H2020-FETFLAG-2018-03 project 820363 OpenSuperQ.

Numerical simulations for the paper were done in QuTiP~\cite{Johansson2012, Johansson2013}. The quantum circuits shown were drawn using quantikz~\cite{Kay2018}.

\end{acknowledgments}


\subsubsection*{Note added} In the final stages of preparing this manuscript, we became aware of patent application PCT/US2019/016047 from Niu et al.~at Google~\cite{Niu2019}, who independently found the idea of the $\mathrm{CCZS} (\theta, \pi, 0)$ gate. In their realization, the only parameter that can be tuned is $\theta$, which is set by the ratio of the amplitudes of the coupling strengths for the two $\ket{11} \leftrightarrow \ket{02}$ transitions involved in the gate.

After this manuscript was submitted for publication, the preprint of Ref.~\cite{Kim2021} appeared on arXiv. In that preprint, it was experimentally demonstrated that simultaneous cross-resonance driving on a chain of three superconducting qubits (as suggested in \secref{sec:Outlook} above) yields a three-qubit iToffoli gate.


\appendix


\section{Simultaneous CZ gates with coupling between qubits 1 and 2}
\label{app:CZ}

In this appendix, we consider the setup from \secref{sec:SetupSimultaneousCZ} with the addition of a direct coupling between qubits 1 and 2. In superconducting circuits, such a coupling can arise due to a small capacitance connecting the two distant qubits or due to a tailored all-to-all coupling in a triangular lattice~\cite{Roth2019}.

For setups where the CZ gates are performed by tuning the states $\ket{1_0 1_j}$ and $\ket{2_0 0_j}$ into resonance, the states $\ket{x01}$ and $\ket{x10}$ with $x = \{ 0, 1, 2\}$ will also become resonant. Then, a direct coupling between qubits 1 and 2 will activate an iSWAP between them. If the CZ gates instead are performed through parametric modulation of tunable couplers between the three qubits, such an iSWAP gate will be activated if there is a frequency component of the modulation in the coupler connecting qubits 1 and 2 that matches the energy difference between the states $\ket{x01}$ and $\ket{x10}$.

We thus have three additions to the diagram in \figpanel{fig:PrincipleCCZS}{b}: a coupling between $\ket{101}$ and $\ket{110}$, transforming the effective $\Lambda$ system in the upper part of the figure into a $\Delta$ system; a coupling between $\ket{201}$ and $\ket{210}$, transforming the effective $V$ system in the lower part of the figure into a $\nabla$ system; and a coupling between $\ket{001}$ and $\ket{010}$. The effect of the last part is simply to change the first term in \eqref{eq:CCZS} from $\ketbra{0}{0}_0 \otimes \mathbb{I}_1 \otimes \mathbb{I}_2$ to $\ketbra{0}{0}_0 \otimes U_{\rm iSWAP} (\beta)$, where
\be
U_{\rm iSWAP} (\beta) = 
\begin{bmatrix}
1 & 0 & 0 & 0 \\
0 & \cos \beta & - i \sin \beta & 0 \\
0 & - i \sin \beta & \cos \beta & 0 \\
0 & 0 & 0 & 1
\end{bmatrix}
\ee
and the angle $\beta = g t$ is determined by the interaction strength $g$ between $\ket{001}$ and $\ket{010}$ and the gate time $t$. In the following, we therefore investigate the dynamics of the effective three-level systems, which will determine the gate time, since we must make sure to return all population to the computational subspace at the end of the gate.

To clarify the dynamics in the full system, we now model the $\Delta$ system separately as in Ref.~\cite{Buckle1986}. For brevity, the states $\ket{101}$, $\ket{200}$, and $\ket{110}$ are renamed $\ket{1}$, $\ket{2}$, and $\ket{3}$, respectively. Denoting the coupling strength on the transition $\ket{i} \leftrightarrow \ket{j}$ by $\alpha_{ij}$ (note that $g = \alpha_{13}$) and assuming that this transition is activated by parametrically modulating the coupling strength with time dependence $\cos \mleft( \nu_{ij} t + \phi_{ij} \mright)$, the Hamiltonian for the $\Delta$ system can be written as
\bea
H_\Delta &=& 2 \alpha_{12} \mleft( \ketbra{1}{2} e^{i \theta_{12}} + \text{H.c.} \mright) \cos \mleft( \nu_{12} t + \phi_{12} \mright) \nn \\
&&+ 2 \alpha_{23} \mleft( \ketbra{2}{3} e^{i \theta_{23}} + \text{H.c.} \mright) \cos \mleft( \nu_{23} t + \phi_{23} \mright) \nn \\
&&+ 2 \alpha_{13} \mleft( \ketbra{1}{3} e^{i \theta_{13}} + \text{H.c.} \mright) \cos \mleft( \nu_{13} t + \phi_{13} \mright) \nn \\
&&+ \sum_i \omega_i \ketbra{i}{i},
\eea
where $\omega_i$ is the energy of state $\ket{i}$. Transforming to a suitable rotating frame using the transformation
\bea
T &=& \exp \mleft[ i \mleft( \nu_{12} t + \phi_{12} - \theta_{12} \mright) \ketbra{1}{1} \mright. \nn \\
&& \mleft. \quad + i \mleft( \nu_{23} t + \phi_{23} + \theta_{23} \mright) \ketbra{3}{3} \mright],
\eea
making the rotating-wave approximation, assuming $\nu_{13} = \nu_{12} - \nu_{23}$, and shifting the zero energy to $\omega_2$, we obtain
\bea
\tilde{H}_\Delta &=& \alpha_{12} \mleft( \ketbra{1}{2} + \ketbra{2}{1} \mright) \nn \\
&&+ \alpha_{23} \mleft( \ketbra{2}{3} + \ketbra{3}{2} \mright) \nn \\
&&+ \alpha_{13} \mleft( \ketbra{1}{3} e^{i \Phi} + \ketbra{3}{1} e^{- i \Phi} \mright) \nn \\
&&+ \Delta_1 \ketbra{1}{1} + \Delta_3 \ketbra{3}{3},
\eea
where
\bea
\Delta_1 &=& \omega_1 - \omega_2 - \nu_{12}, \\
\Delta_3 &=& \omega_3 - \omega_2 - \nu_{23}, \\
\Phi &=& \phi_{12} - \theta_{12} - \phi_{23} - \theta_{23} + \theta_{13} - \phi_{13}.
\eea

Further simplifying the situation by assuming $\Delta_1 = \Delta_3 = 0$ (resonant individual CZ gates) and $\Phi = 0$, introducing the dark and bright states
\bea
\ket{B} &=& \sin \theta \ket{1} + \cos \theta \ket{3}, \\
\ket{D} &=& \cos \theta \ket{1} - \sin \theta \ket{3},
\eea
where $\tan \theta = \alpha_{12} / \alpha_{23}$, and then also assuming that the two individual CZ gates have equal gate strength ($\alpha_{12} = \alpha_{23}$, so $\theta = \pi / 4$), we arrive at
\be
\tilde{H}_\Delta = \sqrt{2} \alpha_{12} \mleft( \ketbra{B}{2} + \ketbra{2}{B} \mright) + \alpha_{13} \mleft( \ketbra{B}{B} - \ketbra{D}{D} \mright).
\ee
In the space spanned by $\ket{B}$, $\ket{2}$, and $\ket{D}$, this Hamiltonian can be written on matrix form as
\bea
\tilde{H}_\Delta &=&
\begin{bmatrix}
\alpha_{13} & \sqrt{2} \alpha_{12} & 0 \\
\sqrt{2} \alpha_{12} & 0 & 0 \\
0 & 0 & - \alpha_{13}
\end{bmatrix}
\nn \\
&=& 
\frac{\alpha_{13}}{2} \mathbb{I} + 
\begin{bmatrix}
\frac{\alpha_{13}}{2} & \sqrt{2} \alpha_{12} & 0 \\
\sqrt{2} \alpha_{12} & - \frac{\alpha_{13}}{2} & 0 \\
0 & 0 & \frac{- 3 \alpha_{13}}{2}
\end{bmatrix}
.
\label{eq:HTildeDeltaMatrixForm}
\eea

From \eqref{eq:HTildeDeltaMatrixForm}, we see that the states $\ket{B}$ and $\ket{2}$ form a two-level system with dynamics governed by the Hamiltonian
\be
H_2 = \frac{\alpha_{13}}{2} \sz + \sqrt{2} \alpha_{12} \sx = \Omega \vec{n} \cdot \vec{\sigma},
\ee
where 
\bea
\Omega &=& \sqrt{\mleft( \frac{\alpha_{13}}{2} \mright)^2 + 2 \alpha_{12}^2}, \\
\vec{n} &=& \mleft( \frac{\sqrt{2} \alpha_{12}}{\Omega}, 0, \frac{\alpha_{13}}{2 \Omega} \mright), \\
\vec{\sigma} &=& \mleft( \sx, \sy, \sz \mright).
\eea
Thus, including the phase, the time evolution of $\ket{B}$ and $\ket{2}$ is given by
\be
e^{- i \frac{\alpha_{13}}{2} t} \mleft[ I \cos \mleft( \Omega t \mright) - \sin \mleft( \Omega t \mright) \vec{n} \cdot \vec{\sigma} \mright],
\label{eq:TimeEvolutionB2}
\ee
while $\ket{D}$ acquires a phase $e^{i \alpha_{13} t}$.

Since we want to eliminate leakage to the $\ket{200} = \ket{2}$ state, which is outside of the computational subspace, we need the off-diagonal elements in \eqref{eq:TimeEvolutionB2} to be zero. This is achieved when $\Omega t = \pi$. Then, the time evolution for $\ket{B}$, $\ket{2}$, and $\ket{D}$ is given by
\be
U =
\begin{bmatrix}
- e^{- i \frac{\alpha_{13}}{2} t} & 0 & 0 \\
0 & - e^{- i \frac{\alpha_{13}}{2} t} & 0 \\
0 & 0 & e^{- i \alpha_{13} t}
\end{bmatrix}
.
\ee
Transforming back to the bare basis of $\ket{1}$, $\ket{2}$, and $\ket{3}$, using that the transformation matrix for going from $\ket{1}$, $\ket{2}$, and $\ket{3}$ to $\ket{B}$, $\ket{2}$, and $\ket{D}$ is
\be 
T =
\begin{bmatrix}
\sin \theta & 0 & \cos \theta \\
0 & 1 & 0 \\
\cos \theta & 0 & - \sin \theta
\end{bmatrix}
,
\ee
we obtain
\begin{widetext}
\be
U_{\rm bare} = 
\begin{bmatrix}
- \frac{e^{- i \frac{\alpha_{13}}{2} t}}{2} \mleft( 1 - e^{- i \frac{\alpha_{13}}{2} t} \mright) & 0 & - \frac{e^{- i \frac{\alpha_{13}}{2} t}}{2} \mleft( 1 + e^{- i \frac{\alpha_{13}}{2} t} \mright) \\
0 & - e^{- i \frac{\alpha_{13}}{2} t} & 0 \\
- \frac{e^{- i \frac{\alpha_{13}}{2} t}}{2} \mleft( 1 + e^{- i \frac{\alpha_{13}}{2} t} \mright) & 0 & - \frac{e^{- i \frac{\alpha_{13}}{2} t}}{2} \mleft( 1 - e^{- i \frac{\alpha_{13}}{2} t} \mright)
\label{eq:UBare}
\end{bmatrix}
.
\ee
\end{widetext}
We see from \eqref{eq:UBare} that a full population transfer between $\ket{101}$ and $\ket{110}$ requires either $t = 4 \pi / \alpha_{13}$ or $\alpha_{13} = 0$. The latter is the case treated in the main text. The former condition shows that a three-qubit gate similar to the CCZS gate in \secref{sec:SimultaneousCZ} can be implemented also when there is an additional nonzero direct coupling $\alpha_{13}$ between qubits 1 and 2.

We finally note that the analysis here for the $\Delta$ system also applies to the $\nabla$ system, if we identify the states $\ket{1}$, $\ket{2}$, and $\ket{3}$ with $\ket{210}$, $\ket{111}$, and $\ket{201}$, respectively. This means that the condition $\Omega t = \pi$ imposed above also ensures that no leakage from the computational subspace takes place in the $\nabla$ system, since the only effect on the state $\ket{111}$ is that it acquires a phase factor $- \exp \mleft( - i \alpha_{13} t / 2 \mright)$, as shown by \eqref{eq:UBare}.


\bibliography{ThreeQubitRefs}

\end{document}